\documentclass[aps,prx,superscriptaddress,notitlepage,nopacs,nofootinbib,amsmath,amstex,amssymb,citeautoscript,longbibliography,twocolumn]{revtex4-2}
\usepackage{physics}
\usepackage{enumitem}
\usepackage{natbib}
\usepackage[english]{babel}
\usepackage{letltxmacro}
\usepackage{latexsym}
\usepackage[T1]{fontenc}
\LetLtxMacro{\ORIGselectlanguage}{\selectlanguage}
\makeatletter
\DeclareRobustCommand{\selectlanguage}[1]{\@ifundefined{alias@\string#1}
    {\ORIGselectlanguage{#1}}
    {\begingroup\edef\x{\endgroup
       \noexpand\ORIGselectlanguage{\@nameuse{alias@#1}}}\x}}
\newcommand{\definelanguagealias}[2]{\@namedef{alias@#1}{#2}}
\makeatother
\definelanguagealias{en}{english}
\definelanguagealias{English}{english}

\usepackage{amsthm}
\usepackage{placeins}
\usepackage{graphicx}
\usepackage{bm}
\usepackage{physics}
\usepackage[dvipsnames]{xcolor}
\usepackage{algorithm}
\usepackage{algpseudocode}
\usepackage{blindtext}
\usepackage{makecell}
\usepackage{comment}
\usepackage[normalem]{ulem}
\makeatletter
\newcommand{\printfnsymbol}[1]{\textsuperscript{\@fnsymbol{#1}}}
\makeatother
\usepackage{hyperref}
\hypersetup{
    unicode=false,
    pdftoolbar=false,
    pdfmenubar=true,
    pdffitwindow=false,
    pdfstartview={FitH},
    pdftitle={},
    pdfauthor={},
    pdfsubject={},
    pdfcreator={},
    pdfproducer={},
    pdfkeywords={},
    pdfnewwindow=true,
    colorlinks=true,
    linkcolor=teal,
    citecolor=teal,
    filecolor=teal,
    urlcolor=teal
}

\begin{document}
\title{
Ergodicity breaking meets criticality in a gauge-theory quantum simulator
}

\author{Ana Hudomal}
\affiliation{Institute of Physics Belgrade, University of Belgrade, 11080 Belgrade, Serbia}

\author{Aiden Daniel}
 \affiliation{School of Physics and Astronomy, University of Leeds, Leeds LS2 9JT, UK}
 
\author{Tiago Santiago do Espirito Santo}
\affiliation{Department of Theoretical and Experimental Physics, Federal University of Rio Grande do Norte, 59078-970 Natal, RN, Brazil}

\author{Milan Kornja\v ca}
\affiliation{QuEra Computing Inc., 1380 Soldiers Field Road, Boston, MA, 02135, USA}

\author{Tommaso Macrì}
\affiliation{QuEra Computing Inc., 1380 Soldiers Field Road, Boston, MA, 02135, USA}

\author{Jad C.~Halimeh}
\affiliation{Department of Physics and Arnold Sommerfeld Center for Theoretical Physics (ASC), Ludwig Maximilian University of Munich, 80333 Munich, Germany}
\affiliation{Max Planck Institute of Quantum Optics, 85748 Garching, Germany}
\affiliation{Munich Center for Quantum Science and Technology (MCQST), 80799 Munich, Germany}
\affiliation{Department of Physics, College of Science, Kyung Hee University, Seoul 02447, Republic of Korea}

\author{Guo-Xian Su}
\affiliation{Department of Physics, Massachusetts Institute of Technology, Cambridge, MA 02139, USA}
\affiliation{MIT-Harvard Center for Ultracold Atoms, Cambridge, MA 02139, USA}

\author{Antun Bala\v z}
\affiliation{Institute of Physics Belgrade, University of Belgrade, 11080 Belgrade, Serbia}

\author{Zlatko Papi\'c}
 \affiliation{School of Physics and Astronomy, University of Leeds, Leeds LS2 9JT, UK}

%\date{\today}
\begin{abstract} 
Recent advances in quantum simulations have opened access to the real-time dynamics of lattice gauge theories, providing a new setting to explore how quantum criticality influences thermalization and ergodicity far from equilibrium.
Using QuEra's programmable Rydberg atom array, we map out the dynamical phase diagram of the spin-1/2 U(1) quantum link model in one spatial dimension by quenching the fermion mass. We reveal a tunable regime of ergodicity breaking due to quantum many-body scars, manifested as long-lived coherent oscillations that persist across a much broader range of parameters than previously observed, including at the equilibrium phase transition point. We further analyze the electron-positron pairs generated during state preparation via the Kibble–Zurek mechanism, which strongly affect the post-quench dynamics. Our results provide new insights into nonthermal dynamics in lattice gauge theories and establish Rydberg atom arrays as a powerful platform for probing the interplay between ergodicity breaking and quantum criticality.
\end{abstract}
\maketitle

\graphicspath{{./Figures/}} %Setting the graphicspath

\section{Introduction}

Gauge theories are the fundamental framework for describing interactions among the elementary constituents of matter in our universe~\cite{aitchison2012gauge,wen2004quantum}. Built on the principle of local gauge invariance, they capture how fermionic matter fields couple to gauge bosons, which forms the foundation of the Standard Model of particle physics~\cite{weinberg2004making}. Their lattice formulations, known as `lattice gauge theories' (LGTs), preserve gauge symmetry on a discrete spacetime grid~\cite{KogutRMP}. This has enabled non-perturbative studies of phenomena such as confinement and chiral symmetry breaking, and computation of many static properties of LGTs via Euclidean-time path-integral methods~\cite{creutz1983monte,gattringer2009quantum,rothe2012lattice,Wiese2013}.

The process of thermalization, which underpins transport and the approach to equilibrium, is key to understanding matter under extreme conditions, from the early universe to heavy-ion collisions~\cite{BergesRMP}. In LGTs, gauge invariance imposes strict local constraints that can dramatically alter relaxation pathways, e.g., due to confinement and string breaking mechanisms. As a result, the interplay between gauge constraints and many-body interactions can give rise to exotic dynamical regimes far from equilibrium. Nevertheless, accessing the real-time dynamics of LGTs remains a major challenge: for example, quantum Monte Carlo techniques suffer from the so-called sign problem~\cite{troyer2005computational}, while tensor-network methods are limited by the rapid growth of entanglement~\cite{banuls2020review}.

Recent progress in programmable quantum simulators has opened a new route to exploring these questions~\cite{georgescu2014quantum,altman2021quantum,halimeh2025quantumsimulationoutofequilibriumdynamics}. Synthetic platforms of trapped ions, superconducting qubits and neutral atoms can natively encode gauge constraints and enable time-resolved measurements of thermalization dynamics in  LGTs~\cite{Yang2020ObservationGaugeInvariance,Zhou2022ThermalizationDynamicsGauge, Su2023, HanYiWang2023,gyawali2025observationdisorderfreelocalizationusing,Mueller2025,than2025phasediagramquantumchromodynamics}.
Notably, Rydberg-atom quantum simulations of a quantum link model (QLM)~\cite{Bernien2017,Bluvstein2021,mark2025observationballisticplasmamemory} in one spatial dimension $(1+1)\mathrm{D}$ -- the minimal model of quantum electrodynamics on a lattice -- revealed an unexpected type of dynamical behavior dubbed `quantum many-body scars' (QMBSs)~\cite{Serbyn2021,MoudgalyaReview,ChandranReview}. This weak form of ergodicity breaking occurs due to gauge-induced kinetic constraints and manifests in special initial conditions that violate the standard paradigm of thermal relaxation in generic quantum many-body systems -- the eigenstate thermalization hypothesis (ETH)~\cite{DeutschETH,SrednickiETH,RigolNature}. 

The existence of QMBSs suggests that gauge-invariance and nonthermal dynamics may intertwine in other unexpected ways. Their interplay may particularly be enhanced at quantum critical points, where correlations become long-ranged due to the closing of an energy gap and entanglement entropy exhibiting universal, logarithmic divergence with system size~\cite{Calabrese2009}. A prominent  example in the $(1+1)\mathrm{D}$ U(1) QLM is the Coleman phase transition, which marks a continuous change from a unique vacuum with vanishing electric flux to a spontaneously parity-broken phase with two degenerate flux vacua~\cite{Coleman1976}. When the system is driven across such a transition at a finite rate, the Kibble–Zurek (KZ) mechanism predicts that critical slowing down prevents adiabatic evolution, leading to the spontaneous formation of topological defects~\cite{Kibble1976,Zurek1985}. In the QLM, these appear as electron-positron pairs whose density reflects the rate of crossing the critical point. While signatures of KZ scaling have been observed in previous Rydberg atom experiments~\cite{Keesling2019}, their impact on QMBS dynamics has not been explored. Moreover, the large critical entanglement near criticality raises the question of whether weak ergodicity breaking can persist in the presence of long-range correlations.

In this work, we use QuEra’s programmable Rydberg-atom quantum simulator~\cite{AquilaWhitepaper} to map out the dynamical phase diagram of a spin-$1/2$ U(1) QLM in $(1+1)\mathrm{D}$ by globally quenching the fermion mass. We identify a broad regime of ergodicity breaking characterized by anomalously slow relaxation and long-lived periodic revivals associated with QMBSs. By ramping the mass across the Coleman transition, we prepare initial states ranging from simple product states to highly entangled ones, demonstrating that scarring dynamics persist even at the quantum critical point, where rapid thermalization was anticipated~\cite{Peng2022,Yao2022criticality,HanYiWang2023}. We further show that the density and internal structure of electron–positron pairs generated via the KZ mechanism play a central role in shaping the post-quench dynamics. Together, these results establish Rydberg atom arrays as a powerful platform for exploring the interplay of gauge constraints, quantum criticality, and ergodicity breaking in LGTs.

The remainder of this paper is organized as follows. In Sec.~\ref{s:model}, we introduce the model, its mapping to the Rydberg atom array, and our experimental setup. In Sec.~\ref{s:phase_diagram} we map out the dynamical phase diagram, finding excellent agreement with numerical predictions. In Sec.~\ref{s:kibble-zurek} we examine the role of electron-positron pairs produced during state preparation on the ergodicity-breaking dynamics after the quench. Finally, Sec.~\ref{s:conclusions} contains our conclusions and outlook, while further details on the mapping between the models, the experimental setup, and the effect of internal structure of electron-positron pairs on the dynamics are given in Appendices and the Supplementary Material (SM)~\cite{SM}.

\begin{figure*}
\includegraphics[width=0.99\textwidth]{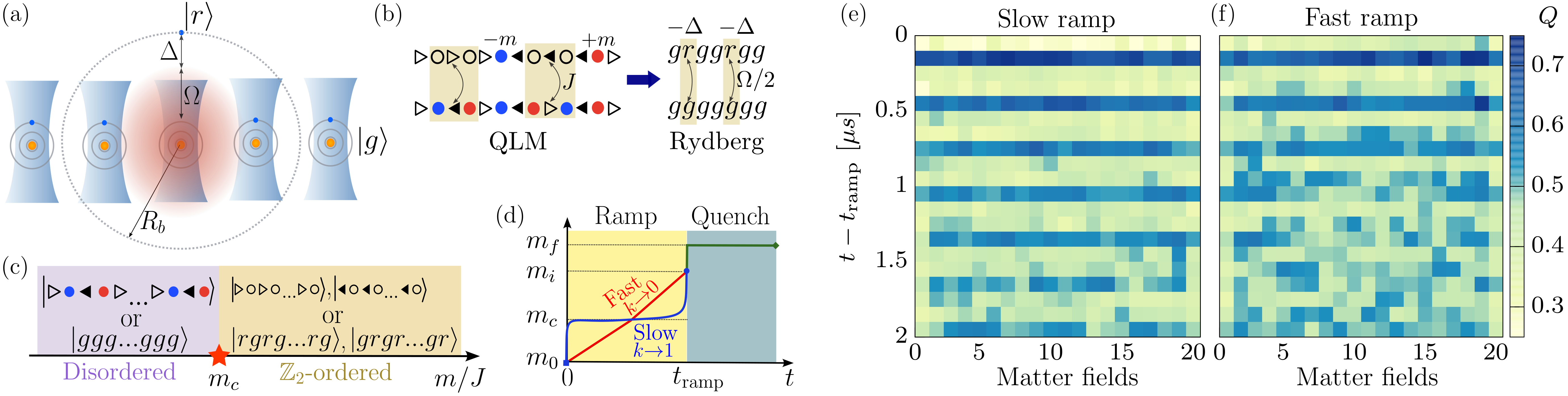}
 \caption{
 \textbf{Simulating the dynamics of a spin-1/2 U(1) QLM in $(1+1)\mathrm{D}$ with Rydberg atoms.}
(a) ${}^{87}\mathrm{Rb}$ atoms are trapped in a 1D tweezer array. The two pseudo-spin states, the ground state $\ket{g}$ and the Rydberg state $\ket{r}$, are coupled by a two-photon process at the Rabi frequency $\Omega$ and detuning $\Delta$.  Nearest-neighbor excitations are suppressed by van der Waals interactions within the Rydberg blockade radius $R_b$, giving rise to the gauge-invariant constraint~\cite{Surace2020}. 
(b) The Rydberg spin states are associated with electric fields in the QLM, with each spin-flip representing a fermion pair production or annihilation process, accompanied by the corresponding change in the direction of electric field (shaded boxes).  This mapping, the details of which are given in Appendix~\ref{a:mapping}, fixes the parameters of the QLM to $J=\Omega/2$ and $m=\Delta/2$, while the average number of electron-positron pairs in the QLM becomes the average density of Rydberg excitations. 
(c) The phase diagram as a function of $m/J$. The representative ground states in the two phases are indicated, in both the Rydberg and QLM representations. The two phases are separated by an Ising phase transition at the critical mass  $m_c\approx 0.65J$. 
(d) Schematic of our two-stage dynamical protocol. The initial state is prepared by ramping from deep in the disordered phase ($m_0 \to -\infty$) into the ordered phase at some $m_i$, potentially crossing the critical point $m_c$. The ramp function $f(x)=x(1-k)/(k-2k\lvert x\rvert +1)$ is a sigmoid with the curvature parameter $k$, see Appendix~\ref{a:experiment}. For present purposes, we only need to distinguish between `fast' (piecewise linear) ramps ($k\to 0$) and `slow' ($k\to 1$) ramps, whose dependencies are sketched in the figure. The second stage of the protocol is the quench where we rapidly change $m_i \to m_f$ and monitor the subsequent dynamics.
(e)-(f) Measurements of the average number of electron-positron pairs after the quench to $m_f=0$. A slow ramp, $k=0.8$ in panel (e), prepares a state close to one of the flux states in panel (c). The subsequent dynamics give rise to persistent oscillations, first observed in Ref.~\cite{Bernien2017}. By contrast, a fast ramp, $k=0$ in panel (f), prepares a state with many electron-positron pairs, which causes a much faster decay of the oscillations in the bulk. The measurements in (e)-(f) were taken at a fixed $J=7.715\, \mathrm{kHz}$.
}
\label{fig:schematic}
 \end{figure*}

\section{Realizing the U(1) quantum link model with Rydberg atoms}\label{s:model}

Our target model, the spin-$1/2$ $\text{U(1)}$ QLM model in $(1+1)\mathrm{D}$, is a minimal model of quantum electrodynamics (QED) on a lattice~\cite{Chandrasekharan1997,Wiese2013}. The Hamiltonian can be written as
\begin{align}
\nonumber \hat{H}_\mathrm{QLM} =& J \sum_{j=1}^{L-1}\hat{\psi}_j^\dagger \hat{U}_{j,j+1}\hat{\psi}_{j+1}+\text{H.c.} 
    \\+&m\sum_{j=1}^L (-1)^j\hat{\psi}_j^\dagger \hat\psi_j
    +\frac{g^2}{2}\sum_{j=1}^{L-1}(\hat E_{j,j+1})^2,
    \label{eq:qlm}
\end{align}
where the fermionic fields $\hat\psi_j, \hat\psi_j^\dagger$ represent matter degrees of freedom on site $j$ with mass $m$, and $L$ is the total number of sites with open boundary conditions. The electric fields $\hat E_{j,j+1}$ reside on the links of the lattice and they are coupled to fermions with strength $J$ via the parallel transporters $\hat U_{j,j+1}$, which obey the commutation relation $[\hat E_{j,j+1},\hat U_{j,j+1}]=\hat U_{j,j+1}$. The last term in Eq.~(\ref{eq:qlm}) represents the electric field energy with gauge coupling $g$.

We adopt the staggered fermion representation~\cite{Kogut1975}, whereby a fermion hopping from an even site to its neighboring odd site creates a positron($\color{red}\bullet$)-electron($\color{blue}\bullet$) pair. The QLM formulation associates the gauge field operators with spin operators, $\hat U_{j,j+1}\rightarrow\hat S_{j,j+1}^+$ and $\hat E_{j,j+1}\rightarrow\hat S_{j,j+1}^z$, and the full QED theory is recovered when the local Hilbert space of spins expands to infinity. For our implementation, the local Hilbert space is truncated to that of a spin-$1/2$, denoting the two local electric field states as $\triangleright$ and $\blacktriangleleft$. As a consequence of this truncation, the last term in Eq.~(\ref{eq:qlm}) is simply an overall constant that can be dropped as it has no effect on the dynamics.

Importantly, the QLM truncation does not compromise the local $\text{U(1)}$ gauge symmetry, generated by
\begin{align}\label{eq:gauss_law}
    \hat{G}_j=\hat \psi_j^\dagger \hat\psi_j -\frac{1-(-1)^j}{2}+\hat{S}^z_{j-1,j}-\hat{S}^z_{j,j+1}\,,
\end{align}
which obeys $[\hat G_j, \hat H_\mathrm{QLM}]=0$ at any site $j$. This constrains the neighboring electric field configurations to the matter field between them, which is a discretized version of the Gauss law. As customary, we will restrict our attention to the physical sector of gauge-invariant states $\lvert\psi\rangle$ which obey the Gauss law condition $\hat{G}_j\lvert\psi\rangle=0$. For our 1D open chain, the Gauss law allows one to completely integrate out matter fields, leaving an effective constrained model for the gauge degrees of freedom only~\cite{Surace2020}.  

Our experiment starts by preparing single ${}^{87}\mathrm{Rb}$ atoms in 1D arrays of equidistant  optical tweezers, with all atoms initialized in the ground state $\ket{g}$, see Fig.~\ref{fig:schematic}(a). The atoms are individually driven into an excited Rydberg state $\ket{r}$ with Rabi frequency $\Omega$. As explained in Appendix~\ref{a:mapping}, this spin-flip dynamics maps to the particle-antiparticle pair production process, which is represented by the kinetic term in Eq.~(\ref{eq:qlm}) upon setting $J=\Omega/2$. The detuning $\Delta$ acts as a chemical potential for Rydberg excitations, which is proportional to the fermion mass in the QLM with $m=\Delta/2$. With this choice of parameters, the gauge-invariant states of the QLM are encoded into Rydberg atom configurations, see Fig.~\ref{fig:schematic}(b) for an example.  Crucially, gauge invariance is ensured by strong van der Waals interactions, which penalize the configurations with neighboring Rydberg excitations, $\ket{\ldots rr \ldots}$. As discussed below, this allows to conveniently certify the accuracy of our simulations and detect any violations of the Gauss law. Unless specified otherwise, all simulations are performed at a fixed $J=7.715\, \mathrm{kHZ}$. 

The phase diagram of the spin-$1/2$ U(1) QLM in $(1+1)\mathrm{D}$ is summarized in Fig.~\ref{fig:schematic}(c). Large negative mass $m\rightarrow-\infty$ energetically favors particle creation, hence the ground state in this limit is the charge-proliferated state with the maximal number of electron-positron pairs, $\lvert{\triangleright}{\textcolor{blue}\bullet}{\blacktriangleleft}{\textcolor{red}\bullet}{\triangleright}{\textcolor{blue}\bullet}...{\blacktriangleleft}{\textcolor{red}\bullet}{\triangleright}{\textcolor{blue}\bullet}{\blacktriangleleft}{\textcolor{red}\bullet}\rangle$. In the opposite limit, $m\rightarrow+\infty$, particle creation is inhibited and the ground state spontaneously breaks $\mathbb{Z}_2$ symmetry by choosing one of two degenerate flux vacua, $\lvert{\triangleright}{\circ}{\triangleright}{\circ} ... {\triangleright}{\circ}\rangle$ and $\lvert{\blacktriangleleft}{\circ}{\blacktriangleleft}{\circ} ... {\blacktriangleleft}{\circ}\rangle$, which contain infinite strings of electric fields and no matter. A quantum phase transition connects the two phases at a critical mass $m_c \approx 0.65J$~\cite{Rico2014}, known as the Coleman phase transition~\cite{Coleman1976}. In a Rydberg atom array, the same phase transition separates the uniform ground state $\ket{\ldots gggg \ldots}$ from the degenerate $\mathbb{Z}_2$-states, $\ket{\ldots rgrg \ldots}$ and $\ket{\ldots grgr \ldots}$.

 Our experimental protocol consists of two stages, sketched in Fig.~\ref{fig:schematic}(d). First, we prepare a pre-quench state $\ket{\psi_i}$ by adiabatically ramping the mass from a large negative value $m_0$ to the initial value $m_i$ over time $t_\mathrm{ramp}=1.5\mathrm{\mu s}$. The ramp follows a sigmoid-type modulation with tunable curvature parameter $k$, sketched in Fig.~\ref{fig:schematic}(d) (see Appendix~\ref{a:experiment} for details). 
In the second stage of the protocol, at $t=t_\mathrm{ramp}$ we abruptly quench $m_i$ to the final value $m_f$ and monitor the subsequent evolution. Generally, we will consider $m_f$ that is different from $m_i$, hence the post-quench dynamics are expected to be complicated because the initial state $\ket{\psi_i}$ is far from any eigenstate of the Hamiltonian at $m_f$. 

We characterize quantum dynamics using the density of matter fields $\hat Q$ in the QLM: 
\begin{equation}
\hat Q = \frac{1}{L-1}
\sum_{j=1}^{L-1}
\left[ \frac{1-(-1)^j}{2} + (-1)^j\,\hat\psi_j^\dagger\hat\psi_j
\right] .
\label{eq:Q}
\end{equation}
With this definition, the uniform flux states in Fig.~\ref{fig:schematic}(c) have $Q=0$, while the charge-proliferated state corresponds to the maximal matter density $Q=1$ for $L\gg 1$. Equivalently, in the Rydberg atom picture, $\hat Q$ counts the number of $\ket{\ldots gg \ldots}$ configurations. Ergodicity breaking is identified by persistent revivals in the charge density $\langle \hat Q(t)\rangle$, which signal a breakdown of conventional thermal relaxation. The latter is sensitive to the  
structure of the initial state, which is controlled by the ramp used to prepare it. In particular, when the fermion mass is ramped across the quantum critical point $m_c$, critical slowing down limits adiabaticity and leads to the formation of domains of uniform electric field separated by particle--antiparticle defects, in accordance with the Kibble--Zurek mechanism~\cite{Kibble1976,Zurek1985}. The typical size and density of these domains depend sensitively on the ramp speed. 

Ramp-induced domain structures play a central role in determining the nature of subsequent post-quench dynamics. A slow ramp with $k=0.8$, Fig.~\ref{fig:schematic}(e), results in longer domains of electric-field strings, $\ldots{\triangleright}{\circ}{\triangleright}{\circ}{\triangleright}{\circ}\ldots$ and $\ldots{\blacktriangleleft}{\circ}{\blacktriangleleft}{\circ} {\blacktriangleleft}{\circ}\ldots$. Such domains exhibit slow dynamics and coherent revivals when quenched to $m_f=0$ -- a hallmark of QMBS dynamics~\cite{Turner2017, wenwei18TDVPscar, Turner2018b, Iadecola2019, Khemani2019, Choi2018, Omiya2022}. By contrast, quench dynamics after a fast ramp with $k=0$, Fig.~\ref{fig:schematic}(f), exhibit much faster dynamics, resulting in a loss of coherence beyond $\approx 1\mathrm{\mu s}$. This behavior is consistent with the expectation that faster ramps nucleate a higher density of domain walls and shorter characteristic length scales~\cite{Keesling2019}. These domain walls destabilize the scarring dynamics of the electric strings connecting them, as we will analyze in more detail in Sec.~\ref{s:kibble-zurek} below.

 \begin{figure*}
\includegraphics[width=0.99\textwidth]{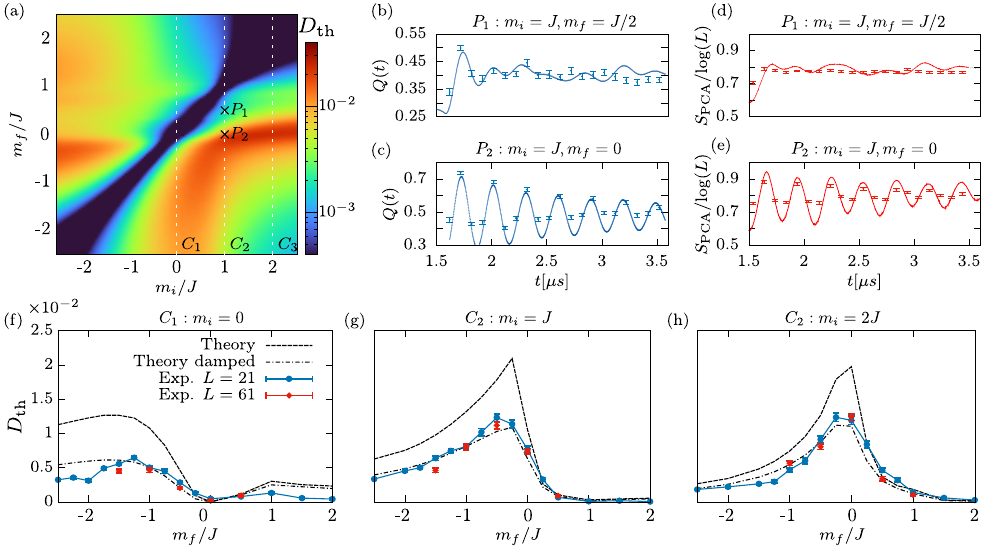}
 \caption{
 \textbf{Dynamical phase diagram of ergodicity breaking in the spin-1/2 U(1) QLM in $(1+1)\mathrm{D}$.}
 (a) Deviation from thermal value, $D_\mathrm{th}$ in Eq.~(\ref{eq:msd}), shown on the color bar, as a function of fermion masses, $m_i$ and $m_f$. Data obtained by exact diagonalization of the Rydberg model [Eq.~(\ref{eq:ryd})] for $L=15$ atoms. Red regions correspond to ergodicity breaking. Dashed lines ($C_1$-$C_3$) are the three cuts that have been measured experimentally, with $C_2$ close to the critical value $m_i\approx J$.   The measured $D_\mathrm{th}$ values at these cuts are plotted in panels (e)-(g), while the dynamics of charge density and PCA entropy at representative points $P_1$ and $P_2$ are shown in panels (b)-(e).  
 (b)-(c) 
 Time series of  QLM charge density $Q(t)$, Eq.~(\ref{eq:Q}), for quenches at $P_1$ ($m_i=J$ and $m_f=J/2$) and $P_2$ ($m_i=J$ and  $m_f=0$), respectively.
 (d)-(e) Time series of the normalized PCA entropy, $S_\mathrm{PCA}/\log(L)$,  for the same quenches as in (b)-(c). 
 In panels (b)-(e), points are experimental data for $L=61$ atoms, while lines are theoretical predictions for $L=21$.
 (e)-(g) Experimental data along the line cuts $C_1$-$C_3$. The data for a smaller system with $L=21$ (blue dots) are seen to be in excellent agreement with the much larger system $L=61$ (red pluses), where we have collected fewer points due to limited resources.
 The error bars in all the plots represent statistical errors. The dashed and dot-dashed black lines show numerical results for $L=21$, without and with exponential damping, respectively.
 }\label{fig:phase_diagram}
 \end{figure*}

\section{Ergodicity-breaking phase diagram}\label{s:phase_diagram}

To systematically explore the existence of ergodicity breaking, we map out the dynamical phase diagram of the QLM in Eq.~(\ref{eq:qlm}) for different combinations of masses $(m_i, m_f)$ in our ramp-quench protocol in Fig.~\ref{fig:schematic}(d). This is a generalization of the original protocol that first observed QMBS in Rydberg atom arrays~\cite{Bernien2017}, where the initial and final mass were kept fixed at $m_i=+\infty$ and $m_f=0$. We quantify ergodicity breaking by measuring the strength of oscillations in the charge density $\hat{Q}$ in Eq.~(\ref{eq:Q})~\cite{Daniel2023}. The mean-square deviation between the instantaneous value of $\hat{Q}$ and its thermal ensemble expectation value $Q_\mathrm{th}$ is given by
\begin{equation}
D_\mathrm{th} \equiv \frac{1}{t_1-t_0}\int_{t_0}^{t_1}\left | \langle \hat{Q}(t) \rangle  - Q_\mathrm{th} \right |^2 \; dt.
\label{eq:msd}
\end{equation}
Here, the observation window is set to $[t_0,t_1]=[1.5, 3.5]~\mu s$ in order to capture the relevant dynamics prior to the onset of decoherence in our experiment. For a thermalizing initial state, the integrand will be close to zero and we expect $D_\mathrm{th}\approx 0$, while a QMBS initial state will lead to a nonzero value of $D_\mathrm{th}$. Since we lack direct experimental access to the thermal Gibbs state, we approximate $Q_\mathrm{th}$ via the time-averaged charge density $Q_\mathrm{av}$, a method validated by our numerical benchmarks~\cite{SM}.

We utilize the slow-ramp protocol shown in Fig.~\ref{fig:schematic}(d) to prepare the initial state, ensuring a controlled approach to the target mass $m_i$ before quenching to $m_f$. As a baseline, in Fig.~\ref{fig:phase_diagram}(a) we first present the $D_\mathrm{th}$ phase diagram obtained by exact diagonalization of the full Rydberg model [Eq.~(\ref{eq:ryd}) in Appendix~\ref{a:mapping}] for $L=15$ atoms. The red regions of high $D_\mathrm{th}$ represent the ergodicity-breaking regimes. As mentioned above, the point ($m_i\rightarrow+\infty$, $m_f=0$) recovers the previously known QMBSs associated with the electric flux states~\cite{Bernien2017}. Remarkably, we see that this point is by no means special: the simultaneous tuning of $m_i$ and $m_f$ gives rise to a continuous line of QMBS revivals and ergodicity breaking across this phase diagram, far beyond the regime of the electric-flux initial states. In particular, the non-ergodic regime extends across the critical mass $m_i = m_c \approx J$, with QMBS effects remaining robust even when the initial state is in the vicinity of the Coleman phase transition. We note that this value of $m_c$ in the Rydberg atom simulator is shifted by the long-range tail of van der Waals interactions compared to its value $m_c \approx 0.65J$ in the QLM~\cite{Rico2014}. Nevertheless, the dynamical phase diagram in Fig.~\ref{fig:phase_diagram}(a) is in good qualitative agreement with that of the QLM in Eq.~(\ref{eq:qlm}), see Ref.~\cite{Daniel2023}. 

Figures~\ref{fig:phase_diagram}(b)-(c) illustrate the typical raw signals from the measurements of the post-quench dynamics for two points, $P_1$ and $P_2$, along the critical line $m_i=J$ in Fig.~\ref{fig:phase_diagram}(a). For each point, we prepare the state at $m_i$, evolve with $m_f$ while measuring the charge density, $Q(t)$, and finally compute $D_\mathrm{th}$ according to Eqs.~\eqref{eq:Q}-\eqref{eq:msd}. The measurements are performed either on a single chain of $L=61$ atoms or on multiple chains of $L=21$ atoms, with the layouts illustrated in Appendix~\ref{a:experiment}. The plots in Fig.~\ref{fig:phase_diagram}(b)-(c) highlight the sensitivity of the frequency, amplitude and decay rate of the QMBS oscillations to the value of  $m_f$. In particular, they distinguish between thermal  [Fig.~\ref{fig:phase_diagram}(b)] and QMBS [Fig.~\ref{fig:phase_diagram}(c)] regimes, capturing the expected oscillation frequency (see SM~\cite{SM} for additional data). We attribute the differences in the oscillation amplitude between the experimental data and numerical simulation to the dephasing caused by experimental errors and environmental influences that are known to occur in this type of experiments~\cite{Bernien2017,Leseleuc2018,AquilaWhitepaper}.

Beyond local observables, we characterize thermalization dynamics using the Principal Component Analysis (PCA) entropy, $S_\mathrm{PCA}$~\cite{Panda2023,Verdel2020,Bhakuni2024} -- a proxy for entanglement entropy, which is challenging to measure~\cite{Islam2015a}.
The PCA entropy is the Shannon entropy defined on a data matrix of size $L\cross n_\mathrm{shots}$, containing the bitstrings resulting from all experimental shots, hence it is readily available in our simulations. Denoting the singular values of this matrix by $s_k$, the PCA entropy is defined as~\cite{Panda2023} 
\begin{equation}\label{eq:PCA}
    S_\mathrm{PCA} = -\sum_k \lambda_k \ln\lambda_k\quad \lambda_k \equiv s_k^2/\sum_l s_l^2\,.
\end{equation}
In Fig.~\ref{fig:phase_diagram}(d)-(e), we plot the measured dynamics of $S_\mathrm{PCA}$, normalized by $\log L$, for the same points as in Figs.~\ref{fig:phase_diagram}(b)-(c). While in the thermal case the PCA entropy quickly reaches saturation, in the QMBS case it exhibits persistent oscillations, signalling a periodic disentangling of the many-body wave function. The experimental revival peaks are well-reproduced by the numerics, although the dips are less pronounced at later times due to decoherence effects mentioned above. Nevertheless, the oscillation frequency of the PCA entropy matches that of the charge density $Q(t)$ in Fig.~\ref{fig:phase_diagram}(c).

Finally, we experimentally probe the phase diagram by taking three representative cuts: $C_{1}$ ($m_{i}=0$) without ramping across the critical point; $C_{2}$ ($m_{i}=J \approx m_{c}$) along the critical line; and $C_{3}$ ($m_{i}=2J$) ramping to the vacuum states with most of the (anti)particles annihilated. The dependence of $D_\mathrm{th}$ on $m_{f}$ for these cuts is shown in Figs.~\ref{fig:phase_diagram}(f)-(h). The cut $C_3$ at $m_i=2J$ serves as a benchmark, since the scarring effects are expected around $m_f=0$, as indeed confirmed by the measurements. The cut $C_{2}$ represents one of our main results: the scarring persists in a highly-entangled initial state at a critical $m_i$. The oscillations in this case are most prominent for a final detuning of $m_{f} = -J/2$, i.e., the oscillation peak is shifted away from $m_{f}=0$ -- compare with the line cut $C_3$. Finally, the $C_{1}$ cut $m_i=0$ is far away from previously known QMBS regimes, with somewhat weaker scarring effects compared to $C_2$ and $C_3$, but with the peak around $m_f \approx -1.2J$ still well above the thermal baseline. 

The experimental results in Figs.~\ref{fig:phase_diagram}(f)-(h) are in good overall agreement with the numerical results obtained by exact diagonalization of the full Rydberg model for $L=21$ atoms (dashed black lines). For all three cuts, the shapes of the numerical curves closely follow the experimental data, albeit with systematically larger values of $D_\mathrm{th}$. This is unsurprising as our numerics assume a closed system. At a phenomenological level, we can account for environmental decoherence effects by introducing an exponential damping of the oscillations, $Q(t) e^{-\alpha t}$, see SM for details~\cite{SM}. The damping coefficient $\alpha=0.5\, \mathrm{kHz}$ was determined empirically and the same value is used for all data points. The resulting damped numerical curves, shown by the dot-dashed black lines in Figs.~\ref{fig:phase_diagram}(f)-(h), are in excellent agreement with the experimental data.

Finally, to demonstrate the stability of our results, we have also collected data for a larger chain of $L=61$ atoms at selected $m_{f}$ values. These measurements [red crosses in Figs.~\ref{fig:phase_diagram}(f)-(h)] are seen to align well with the $L=21$ results, indicating that the $D_\mathrm{th}$ curves do not move significantly with increasing system size and our conclusions are expected to hold in the thermodynamic limit on timescales probed in Fig.~\ref{fig:phase_diagram}.

\begin{figure*}[tbh!]
 \includegraphics[width=0.99\textwidth]{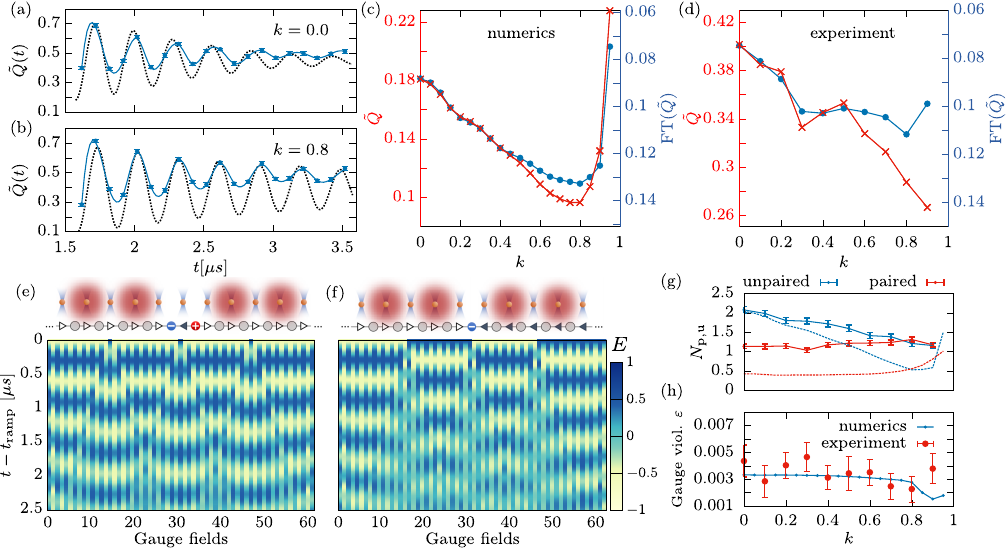}
 \caption{
 \textbf{Impact of electron-positron pairs on the nonergodic dynamics in the spin-1/2 U(1) QLM in $(1+1)\mathrm{D}$.}
 (a)-(b) The post-quench dynamics of total matter field density, $\tilde{Q}$ in Eq.~(\ref{eq:n_d}),  for a fast ramp with $k=0$ (a) versus slow ramp with $k=0.8$ (b). Dots are experimental results, solid line is their interpolation, dashed line is the numerical simulation of the Rydberg model [Eq.~(\ref{eq:ryd})].
 (c)-(d) $\tilde Q$ in the initial state (red crosses, left $y$-axis) compared against the first peak of the Fourier transform of $\tilde Q(t)$ (blue circles, right $y$-axis), as a function of ramp curvature $k$. Close agreement between the two for $k\lesssim 0.6$ demonstrates that the density of electron-positron pairs in the initial state governs the decay of QMBS oscillations. Panel (c) is numerics, (d) is experimental data. 
 (e)-(f) Beyond the sheer number of electron-positron pairs, different \emph{types} of particle-string domains can have a strong impact on QMBS dynamics. This is illustrated by the numerical simulations of the dynamics for `even' (e) vs. `odd' (f) type of particle-string defects in the initial state, schematically shown on top of each plot. 
 (g) The measured number of paired (e.g., $\textcolor{red}{\bullet}\textcolor{blue}{\bullet}\textcolor{red}{\bullet}\textcolor{blue}{\bullet}$) and unpaired (e.g., $\textcolor{red}{\bullet}\textcolor{blue}{\bullet}\textcolor{red}{\bullet}$) particle strings, which serve as domain walls between vacuum domains with uniform electric field. Dashed lines are theoretical predictions.
 (h) Density of gauge violations $\varepsilon$ after the ramp and before the quench.
 In all the panels, the parameters $(m_i=2J, m_f=0)$ correspond to the point with the most robust QMBS signatures along the $C_3$ line in Fig.~\ref{fig:phase_diagram}.  
 The system size used in all panels except (e) and (f) for both the experiment and the numerics is $L=21$. Panels (e) and (f) are numerics for $L=61$ atoms using time-dependent variational principle algorithm in TenPy~\cite{tenpy}, with a maximum bond dimension $1000$.
}
 \label{fig:kz_exp}
 \end{figure*} 
 
\section{Interplay of scarring and domain-wall dynamics}\label{s:kibble-zurek}

When ramping from the charge-proliferated state into the vacuum state, Coleman's phase transition is crossed, leading to the nucleation of domain walls according to the Kibble-Zurek (KZ) mechanism \cite{Kibble1976,Zurek1985}. In the LGT context, these domain walls correspond to residual (anti)particles that separate distinct domains of the electric field.  The scaling of the domain-wall density with ramp speed has been extensively studied \cite{delCampo2014,Navon2015,Chepiga2021}. In particular, the associated critical exponent has been measured in Rydberg atom arrays \cite{Keesling2019},  and numerical studies have found the exponent to be highly sensitive to the details of the ramp trajectory~\cite{Garcia2024b}. Furthermore, a breakdown of KZ scaling was numerically observed when the change rate of the control parameter is too fast or too slow~\cite{Zeng2023,Rao2025}.

Here we investigate the impact of domain walls on the subsequent ergodicity-breaking dynamics. We focus on the quench regime where QMBSs are most pronounced, ramping to the pre-quench mass of $m_{i}=2J$ and quenching to $m_{f}=0$, i.e., the peak of $D_\mathrm{th}$ in Fig.~\ref{fig:phase_diagram}(h). We use the same ramp-quench protocol as in  Fig.~\ref{fig:schematic}(d) and tune the curvature parameter $k$ of the sigmoid ramp, which continuously varies the adiabaticity of state preparation. For all values of $k$, the ramp function is approximately linear in the vicinity of the transition, and the pre-quench state consists of electric flux domains separated by (anti)particles. To quantify the impact of these pre-quench domain walls, we measure the \emph{total} matter-field density
\begin{equation}\label{eq:n_d}
    \tilde{Q} \equiv Q + \langle \hat{\varepsilon}\rangle, \quad \hat{\varepsilon} = \frac{1}{L-1}
\sum_{j=1}^{L-1}
\langle \hat G_j^2 \rangle, 
\end{equation}
which includes the possible violations of the Gauss law in Eq.~(\ref{eq:gauss_law}) given by $\hat{\varepsilon}$.  The gauge violations correspond to neighboring Rydberg excitations, $\ket{\ldots rr\ldots}$, which are rare since they are suppressed by the Rydberg blockade. 

The typical dynamics of $\tilde{Q}(t)$, for illustrative fast and slow ramps, are shown in Fig.~\ref{fig:kz_exp}(a)-(b). Increasing $k$ (slower ramp) reduces the initial density of particle-antiparticle pairs, which leads to longer coherent oscillations. This improvement persists up to $k \approx 0.85$, beyond which the ramp becomes too steep at the beginning and at the end, causing non-adiabaticity and resurgence in domain wall production.

Intuitively, the density of particle-antiparticle pairs is expected to impact the amplitude of the QMBS oscillations. We quantify this by evaluating the Fourier transform of $\tilde{Q}(t)$ and extracting the height of the first peak, which is then compared with the density of domain walls in the initial state (prior to the quench) as a function of $k$. The numerical and experimental results are shown in Figs.~\ref{fig:kz_exp}(c) and (d), respectively. In both cases, we find excellent agreement between the two for $k\lesssim 0.5$. Some discrepancy occurs at larger values of $k$, when the dominant type of defects starts to change, as we discuss next.

Crucially, due to the constrained Hilbert space, the post-quench dynamics are governed not merely by the total density of domain-wall particles but also by their microscopic structure. We identify two distinct classes of domain walls based on the alignment of the electric-flux domains. The first class, even-length domain walls, corresponds to localized particle-antiparticle pairs ($\ldots{\triangleright}{\textcolor{blue}\bullet}{\blacktriangleleft}{ \textcolor{red}\bullet}{\triangleright}\ldots$ or $\ldots{\blacktriangleleft}{\textcolor{red}\bullet}{ \triangleright}{ \textcolor{blue}\bullet}{\blacktriangleleft}\ldots$) situated between electric field domains that are aligned in the same direction, see Fig.~\ref{fig:kz_exp}(e). These domain walls are dynamically active and, since we work in the deconfined regime of the QLM~\cite{Desaules2023confinement}, the spreading of these particles is unbounded. These pairs oscillate out of phase with the remaining electric flux domains; the particle and the antiparticle move ballistically away from each other, creating expanding domains between the pairs. However, the expanding domains oscillate at the same frequency as the original domains, and therefore coherent oscillations are sustained. The expansion of these domains may be suppressed by periodically driving the detuning, as previously demonstrated for the pure flux states~\cite{Hudomal2022Driven}.

In contrast, the odd-length domain walls contain a single unpaired (anti)particle ($\ldots{\triangleright}{\textcolor{blue}\bullet}{ \blacktriangleleft}\ldots$ or $\ldots{\blacktriangleleft}{\textcolor{red}\bullet}{\triangleright}\ldots$). They separate two opposite domains of the electric field, making them topologically distinct from the type discussed above. As illustrated in Fig.~\ref{fig:kz_exp}(f), these (anti)particles `melt away' locally as the system evolves. However, this local thermalization does not succeed in producing a homogenized vacuum order, as the electric field domains on either side of the domain wall remain largely pinned to their original, anti-aligned direction. Healing such domain walls would require a macroscopic reversal of the electric flux, a process that is energetically costly and dynamically suppressed. Consequently, these `melting' domain walls act as dissipative barriers that fracture the system into finite-sized segments, accelerating the decay of a coherent QMBS signal.

The internal structure of the domain walls evolves distinctly as the ramp adiabaticity is tuned, as summarized in Fig.~\ref{fig:kz_exp}(g). In the fast-ramp regime ($k \rightarrow 0$), the domain wall population is dominated by odd-length clusters, i.e., clusters containing unpaired (anti)particles. However, as the ramp curvature is increased toward the adiabatic limit ($k \rightarrow 1$), we observe a crossover at $k \approx 0.6$ in the domain wall statistics. The number of single (anti)particles significantly decreases, while the relative fraction of pairs monotonically increases. Therefore, as the ramp becomes more adiabatic, the statistical shift of the domain wall structure, together with the reduction of the total number of domain walls, helps protect the coherence of QMBS oscillations. 
The numerical results exhibit the same qualitative behaviour, with a decreasing number of unpaired and an approximately constant number of paired (anti)particles. The remaining quantitative discrepancies can likely be attributed to readout errors in the experiment, which can artificially truncate an (anti)particle string and thereby change its parity.
Furthermore, we find that the number of gauge violations decreases slightly when using a slower ramp with higher $k$, meaning that such ramp parameters also lead to a better agreement with the exact QLM, see Fig.~\ref{fig:kz_exp}(h).

In summary, our combined experimental and numerical analysis shows that the structure of ramp-induced domain walls dictates the decay of QMBS oscillations through a complex interplay of defect density and internal structure. Adjacent particle--antiparticle pairs are capable of dynamical healing, effectively reconnecting aligned electric-flux domains and sustaining coherent oscillations, whereas isolated (anti)particles fragment the system into finite domains where decoherence is accelerated. Experimental analysis of the domain-wall statistics demonstrates that slower ramp trajectories suppress the total defect density and simultaneously decrease the statistical fraction of non-healable single-particle domain walls, leading to better preservation of coherence as the ramp becomes more adiabatic. Taken together, these results establish a direct connection between KZ state preparation and the robustness of nonergodic post-quench dynamics in the U(1) QLM.

\FloatBarrier

\section{Conclusions and outlook}\label{s:conclusions}

By exploiting the mapping between constrained Rydberg excitations and gauge-invariant degrees of freedom, we used a programmable Rydberg-atom array to study quench dynamics of the spin-$1/2$ U(1) QLM in $(1+1)\mathrm{D}$ over a wide range of fermion masses. We found persistent ergodicity-breaking signatures throughout the dynamical phase diagram, including near the Coleman quantum phase transition. This shows that quantum criticality does not generically destabilize weak ergodicity breaking, but can coexist with, and even enhance, the dynamics of nonthermal states in gauge theories. Crucially, our results reveal a direct connection between domain formation during state preparation and the decay time scale of QMBS oscillations, linking KZ scaling to the stability of nonthermal dynamics. Together, these findings establish programmable Rydberg-atom arrays as a powerful platform for uncovering microscopic mechanisms that control ergodicity breaking in LGTs far from equilibrium.

Our work opens many future directions. 
The mechanism behind robust QMBS revivals from the critical, highly-entangled ground state remains to be understood. Ref.~\cite{Daniel2023} pointed out that the post-quench Hamiltonian in this case exhibits nearly dispersionless quasiparticle excitations at low energies. However, the link between non-interacting quasiparticles at low energies and QMBSs at much higher energies is not transparent, nor is it clear why this occurs at the particular value of $m_f$. It would be interesting to design new experimental probes that could probe the nature of interactions between quasiparticles in this setup. Another natural extension of our setup are the higher-dimensional Rydberg lattices with richer lattice connectivity and plaquette-type gauge constraints. While scar dynamics in two-dimensional Rydberg arrays have so far been observed only on bipartite lattices~\cite{Bluvstein2021}, it would be interesting to explore whether our ramp-and-quench protocol can stabilize ergodicity breaking on non-bipartite lattices, where geometric frustration and gauge constraints compete. Another promising direction is non-Abelian gauge theories, whose richer Hilbert-space structure is expected to significantly modify the interplay between criticality, kinetic constraints, and ergodicity breaking. It would also be illuminating to combine quantum-scar protocols with periodic (Floquet) driving or engineered dissipation, to test whether nonthermal oscillations can be stabilized or synchronized by external control~\cite{Maskara2021,Hudomal2022Driven}. On the theoretical side, our results highlight the need for an overarching description of the interplay between disjoint $\mathbb{Z}_2$ (electric flux) domains, KZ critical scaling and stability of QMBS dynamics.

\emph{Note added:} During the final stages of this work, a related preprint appeared demonstrating that defect statistics in slow KZ ramps may be strongly affected by non-critical coarsening dynamics, leading to correlated defect formation in a Rydberg-atom quantum simulator~\cite{balewski2025observationanomalystatisticskibblezurek}. These results are consistent with our observation that slow ramps can lead to the formation of aligned electric-field domains.

\begingroup
\appendix

%\FloatBarrier

\begin{acknowledgments}
A.H.~and A.B.~acknowledge funding provided by the Institute of Physics Belgrade, through the grant by the Ministry of Science, Technological Development, and Innovations of the Republic of Serbia. 
A.H. acknowledges financial support from the L’Oréal-UNESCO ``For Women in Science'' National Award.
Part of the numerical simulations were performed at the Scientific Computing Laboratory, National Center of Excellence for the Study of Complex Systems, Institute of Physics Belgrade.
Part of the numerical simulations were performed at the High Performance Computing Center of the Federal University of Rio Grande do Norte (NPAD/UFRN).
J.C.H.~acknowledges funding by the Max Planck Society, the Deutsche Forschungsgemeinschaft (DFG, German Research Foundation) under Germany’s Excellence Strategy – EXC-2111 – 390814868, and the European Research Council (ERC) under the European Union’s Horizon Europe research and innovation program (Grant Agreement No.~101165667)—ERC Starting Grant QuSiGauge. Views and opinions expressed are, however, those of the author(s) only and do not necessarily reflect those of the European Union or the European Research Council Executive Agency. Neither the European Union nor the granting authority can be held responsible for them. This work is part of the Quantum Computing for High-Energy Physics (QC4HEP) working group.
Z.P. acknowledges support by the Leverhulme Trust Research Leadership Award RL-2019-015 and EPSRC Grants EP/Z533634/1, UKRI1337. This research was supported in part by grant NSF PHY-2309135 to the Kavli Institute for Theoretical Physics (KITP). 
\end{acknowledgments}

\section{Mapping between QLM and Rydberg models}\label{a:mapping}

In this work, we employ QuEra's Aquila device~\cite{AquilaWhitepaper} to simulate the dynamics of the spin-$1/2$ $\text{U(1)}$ QLM in Eq.~(\ref{eq:qlm}). The many-body Hamiltonian of a 1D Rydberg atom tweezer array can be written as
\begin{equation}
    \hat{H}_\mathrm{Ryd}=\frac{\Omega}{2} \sum_{j=1}^L \hat{X}_j -\Delta \sum_{j=1}^L \hat{n}_j + \sum_{i<j}^L V_{ij}\hat{n}_i\hat{n}_j\,,
    \label{eq:ryd}
\end{equation}
where $\hat{X} \equiv \lvert g\rangle\langle r\rvert+\lvert r\rangle\langle g\rvert$ is the Pauli-$x$ operator, which encodes Rabi oscillations with the frequency $\Omega$ [Fig.~\ref{fig:schematic}(a)], $\hat{n} \equiv \lvert r\rangle\langle r\rvert$ counts the Rydberg excitations, $\Delta$ is the detuning, and $V_{ij}=C_6/r^6_{ij}$ is the van der Waals interaction between two atoms in the excited states at a distance $r_{ij}$. The total number of Rydberg excitations will be denoted by $N=\sum_i \langle\hat n_i\rangle$ and the average excitation density by $n=N/L$, where $L$ is the number of atoms in the chain. 

When interactions are sufficiently strong, $V_{i,i+1} \gg \Omega, \Delta \gg V_{i,i+2}$, neighboring excitations $\ldots rr \ldots$ become energetically unfavorable. This defines the Rydberg blockade radius $R_b$, where the interaction strength is equal to the Rabi frequency $C_6/R^6_b=\Omega$. When setting the blockade radius to $a<R_b<2a$, the model in Eq.~(\ref{eq:ryd}) can be mapped to the spin-$1/2$ QLM in Eq.~\eqref{eq:qlm}, with the parameters set to $J=\Omega/2$ and $m=\Delta/2$~\cite{Surace2020}, see Table~\ref{tab:mapping}.

Using the staggered-fermion representation, electrons reside on odd matter sites, while positrons reside on even matter sites. Gauss's law is enforced by the Rydberg blockade that only allows one Rydberg excitation for adjacent atoms, which allows us to integrate out matter fields and represent gauge fields using Rydberg configurations~\cite{Surace2020}. Specifically, $\ket{rg}$, $\ket{gg}$, $\ket{gr}$, are mapped to ${\blacktriangleleft}{\circ}{\blacktriangleleft}$, ${\blacktriangleleft}{\color{red}\bullet}{\triangleright}$ and ${\triangleright}{\circ}{\triangleright}$ for odd-even sites, and ${\triangleright}{\circ}{\triangleright}$, ${\triangleright}{\color{blue}\bullet}{\blacktriangleleft}$, ${\blacktriangleleft}{\circ}{\blacktriangleleft}$ and for even-odd sites, while the gauge violation $\ket{rr}$ is energetically suppressed. For the balance between gauge violation and the $1/r^6$ tail of the Rydberg interaction, we have chosen the optimal Rydberg blockade radius $R_b=1.4 a$~\cite{SM}. 

Finally, utilizing the mapping of states in Table~\ref{tab:mapping}, QLM observables introduced in the main text can be compactly expressed in Rydberg-atom language using the local density of excitations $\hat n_j$. For example, matter density $\hat Q$ in Eq.~(\ref{eq:Q}) assumes the form $\hat Q = 1/(L-1) \sum_{j=1}^{L-1} (1-\hat n_j)(1-\hat n_{j+1})$, while the gauge violation $\hat\varepsilon$ in Eq.~(\ref{eq:n_d}) becomes $\hat\varepsilon = 1/(L-1)\sum_j \hat n_j \hat n_{j+1}$. Thus, by counting the number of $\ldots gg\ldots$ and $\ldots rr \ldots$ pairs in measured bitstrings, we characterize the evolution of particle-antiparticle pairs in the QLM and the violation of the Gauss law. 

\begin{center}
\begin{table}[t]
\begin{tabular}{|c|c|}  
 \hline
 \multicolumn{2}{|c|}{\textbf{Mapping}} \\
 \hline\hline
 \textbf{QLM} & \textbf{Rydberg}  \\ 
 \hline\hline
 Gauge fields & Atoms\\
 \hline
 & $r$ (odd sites)\\
 ${\triangleright}$ & or \\
 & $g$ (even sites)\\
 \hline
  & $g$ (odd sites)\\
 ${\blacktriangleleft}$ & or\\
& $r$ (even sites)\\
 \hline\hline
 Matter fields & Links between atoms\\
 \hline
 positron ($e^+$) $\textcolor{red}{\bullet}$ & $gg$ (even links)\\
 electron ($e^-$) $\textcolor{blue}{\bullet}$ & $gg$ (odd links)\\
 empty $\circ$ & $rg$ or $gr$\\
 \hline\hline
 \multicolumn{2}{|c|}{Parameters} \\
\hline
 energy scale $J$ & Rabi frequency $\Omega$/2\\
 fermionic mass $m$ & Detuning $\Delta/2$\\
 \hline\hline
 \multicolumn{2}{|c|}{Special states} \\
 \hline
 \makecell{Extreme vacua \\ or flux \& anti-flux states \\ $\lvert{\triangleright}{\circ}{\triangleright}{\circ}...{\triangleright}{\circ}\rangle$ and $\lvert{\blacktriangleleft}{\circ}{\blacktriangleleft}{\circ}...{\blacktriangleleft}{\circ}\rangle$}& {\makecell{$\mathbb{Z}_2$ states \\ $\lvert {r}{g}{r}{g}...{r}{g}{\rangle}$ and $\lvert {g}{r}{g}{r}...{g}{r}\rangle$}}  \\
 \hline
 \makecell{Charge-proliferated state \\ $\lvert{\triangleright}{\textcolor{blue}\bullet}{\blacktriangleleft}{\textcolor{red}\bullet}{\triangleright}{\textcolor{blue}\bullet}...{\blacktriangleleft}{\textcolor{red}\bullet}{\triangleright}{\textcolor{blue}\bullet}{\blacktriangleleft}{\textcolor{red}\bullet}\rangle$} & {\makecell{Polarized state \\$\lvert {g}{g}{g}...{g}{g}{g}\rangle$}}  \\
 \hline\hline
 \multicolumn{2}{|c|}{Defects} \\
 \hline
 {\makecell{Gauge violation  \\${\blacktriangleleft}\textcolor{blue}\bullet{\triangleright}$ or ${\triangleright}\textcolor{red}\bullet{\blacktriangleleft}$}} & {\makecell{Rydberg blockade violation\\  $rr$}} \\
 \hline
 {\makecell{Separated $e^{-}{-}e^{+}$ pair \\ ${\triangleright}{\textcolor{blue}\bullet}{\blacktriangleleft}{\circ}...{\circ}{\blacktriangleleft}{\textcolor{red}\bullet}{\triangleright}$}} &  {\makecell{Even-length polarized domain  \\ ${r}{g}{g}{r}$}} \\
 \hline
 {\makecell{Adjacent $e^{-}{-}e^{+}$ pair \\ ${\triangleright}{\textcolor{blue}\bullet}{\blacktriangleleft}{\textcolor{red}\bullet}{\triangleright}$}} & {\makecell{Odd-length polarized domain \\ ${r}{g}{g}{g}{r}$}}\\
 \hline
 \end{tabular}
 \centering
 \caption{Mapping of the spin-$1/2$ QLM from Eq.~\eqref{eq:qlm} onto the Rydberg atom array described by Eq.~\eqref{eq:ryd}.
 }
 \label{tab:mapping}
 \end{table}
\end{center}

%\FloatBarrier
\section{Experimental setup}\label{a:experiment}
 
Our experiment starts by loading single atoms in optical tweezers and rearranging them into 1D  arrays, with all atoms initialized in their electronic ground state $\ket{g}=\ket{5S_{1/2}}$. We drive the atoms to the excited Rydberg state $\ket{r}=\ket{70S_{1/2}}$ 
by the two-photon scheme with counterpropagating lasers at 420~nm and 1013~nm, 
which yields the Rabi frequency $\Omega$. The detuning $\Delta$ gives rise to the chemical potential for Rydberg excitations. The Rabi frequency $\Omega(t)$ and detuning $\Delta(t)$ are tuned by power and detuning of Rydberg excitation lasers, within the technical specifications listed in the SM~\cite{SM}. 

We set the initial detuning to $\Delta_0=-8.1\Omega$ and then increase it over a time of $t_\mathrm{ramp}=1.5~\mathrm{\mu s}$ using a ramp based on the sigmoid-like function, defined in the caption to Fig.~\ref{fig:schematic}(d). This slowly brings the detuning
from a large negative value up to the desired value $\Delta_i$, see Fig.~\ref{fig:ramp}(a). 
If we do not cross the phase transition, $\Delta_i\leq\Delta_c$, we use the half-sigmoid in the range $x\in[-1,0]$, otherwise for $\Delta_i>\Delta_c$ we use the full sigmoid with $x\in[-1,1]$.
The function $f(x)$ is shifted and rescaled such that the initial point $x=-1$ corresponds to $(t,\Delta)=(0,\Delta_0)$ and the final point ($x=0$ or $x=1$) to $(t,\Delta)=(1.5\mu s,\Delta_i)$. If we cross the transition, $\Delta_i>\Delta_c$, the case depicted in Fig.~\ref{fig:ramp}(b), the inflection point %$\mathrm{d}f(x)/\mathrm{d}x=0$ 
is at $(t,\Delta)=(0.75\mu s,\Delta_c)$, ensuring that the flattest part of the ramp is around the critical detuning.

At $t=1.5\mu s$, the detuning is quenched to $\Delta_f$ and the system is left to evolve, with measurements performed at $20$ different time points $t_m$ in the interval $[1.5\mu s,3.5\mu s]$. The functional form of $\Delta(t)$ is summarized as:
\begin{align}\label{eq:delta}
    \Delta(t)=\begin{cases}
        \Delta_0, \quad\quad\quad t=0\\
        \Delta_\mathrm{sig}(t), \quad\ \  t< t_\mathrm{ramp}\\
        \Delta_i, \quad\quad\quad\  t = t_\mathrm{ramp}\\
        \Delta_f, \quad\quad\quad t_\mathrm{ramp}<t\leq t_m.
    \end{cases}
\end{align}

\begin{figure}[t]
    \centering
\includegraphics[width=0.9\linewidth]{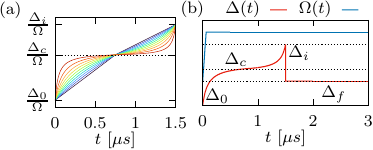}
    \caption{{\bf Ramp protocol.} (a) Plot of the sigmoid part of the ramp from Fig.~\ref{fig:schematic}(d), for different values of $k$. (b) The complete protocol, with the simultaneous modulation of both $\Delta$ and $\Omega$, according to Eqs.~(\ref{eq:delta})-(\ref{eq:omega}).}
    \label{fig:ramp}
\end{figure}

The functional form of the Rabi frequency is as follows:
\begin{align}\label{eq:omega}
    \Omega(t)=\begin{cases}
        t\left(\frac{\mathrm{d}\Omega}{\mathrm{d}t}\right)_\mathrm{max}, \quad\quad\quad t\leq t_{r,\Omega}\\
        \Omega, \quad\quad\quad\quad\quad\quad\ \  t_{r,\Omega}<t\leq t_m-t_{r,\Omega}\\
        \Omega-t\left(\frac{\mathrm{d}\Omega}{\mathrm{d}t}\right)_\mathrm{max}, \quad t_m-t_{r,\Omega}<t\leq t_m,
    \end{cases}
\end{align}
with $\Omega(t)$ linearly increasing with the maximal slew rate from 0
to a constant value $\Omega$ during the ramp time $t_{r,\Omega}=0.06~\mathrm{\mu s}$, see Fig.~\ref{fig:ramp}(b). In the end, $\Omega(t)$ is linearly reduced to 0 before measurement time $t_m$. The constant value Rabi frequency was set to $\Omega=15.43~\mathrm{rad}/\mathrm{\mu s}$, corresponding to the Rydberg blockade radius of $R_b=8.6~\mathrm{\mu m}$. In order to obtain the desired ratio $R_b=1.4a$, which was numerically shown to provide the best agreement of the phase diagram with that of the desired effective model, the interatomic distance was chosen to be $a=6~\mathrm{\mu m}$. 

 \begin{figure}[t]
\includegraphics[width=0.9\linewidth]{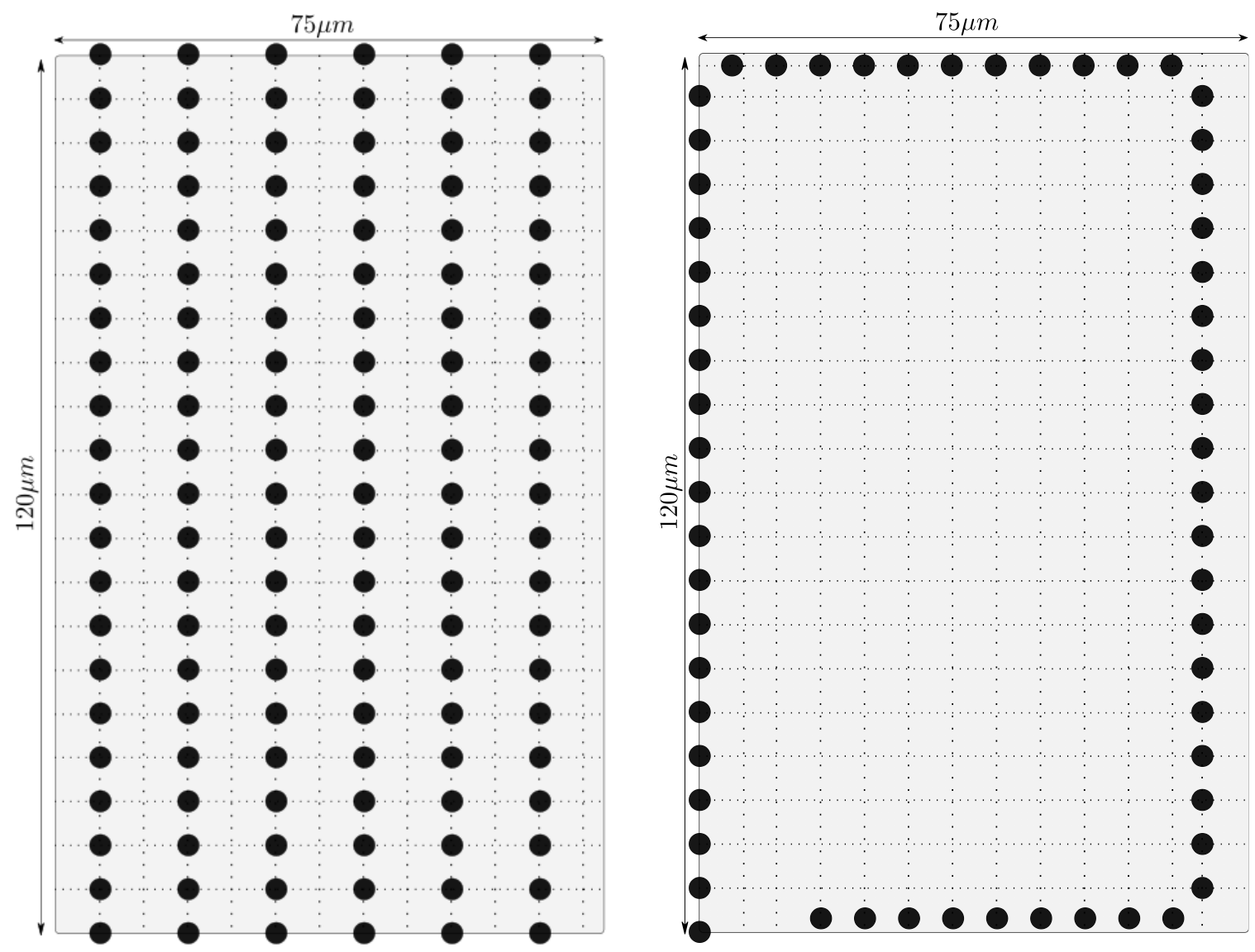}  
 \caption{
 \textbf{Spatial arrangement of atoms.}
 Left panel shows six parallel chains of 21 atoms, right panel is one long chain of 61 atoms looping around the available physical space.
 }\label{fig:atoms}
 \end{figure}

The atoms were arranged in equidistant 1D chains, either in
six parallel linear chains of 21 atoms, or as one longer chain of 61 atoms spiralling around the available $75~\mathrm{\mu m}\times120~\mathrm{\mu m}$ physical space, see Fig.~\ref{fig:atoms}. The short-chain setup requires less experimental shots (approximately 40 shots needed per data point) and can be directly compared to exact diagonalization. The long-chain setup has a higher error rate and requires more experimental shots (around 200), but provides results for larger systems that cannot be reached by exact diagonalization. We therefore used the short-chain configuration for the majority of experimental simulations, while confirming the applicability of our results in larger systems by testing the long-chain configuration for a smaller number of experimental data points. 

If the ramp is slow enough in the vicinity of the critical point, we are able to successfully prepare the desired initial state, whose behavior is then observed over the following $2~\mathrm{\mu s}$ after quenching to $\Delta_f$. At each time step, approximately 200 experimental shots are performed and quantities such as the number of atoms in the excited Rydberg state are extracted from the resulting data. 

Experimental output is obtained in the form of a series of bitstrings, with zeroes and ones representing $\ket{g}$ and $\ket{r}$ states, respectively. The total number of bitstrings is equal to the number of successful experimental shots, as shots with detected atom loss are automatically discarded. The number of successful shots was typically around $200$ for $21$-atom and $120$ for $61$-atom chains.
In the SM~\cite{SM}, we provide additional details about the statistics of experimental bitstrings.

In order to reduce the effect of readout errors, we also perform error mitigation techniques~\cite{Bravyi2021}. We take into account that the known probabilities of erroneously measuring the state of an atom as ground instead of excited is $p_{r\rightarrow g}=8\%$, and excited instead of ground $p_{g\rightarrow r}=1\%$~\cite{SM}. This procedure results in improved agreement of experimental measurements of $Q(t)$ and $D_\mathrm{th}$ with the numerical simulations. We use a modified error mitigation procedure for the gauge violation density $\varepsilon$, as explained in the SM~\cite{SM}. The technique from Ref.~\cite{Bravyi2021}, however, is not suitable for the computation of $S_\mathrm{PCA}$, for which we have used the raw experimental bitstrings.
\endgroup
\FloatBarrier

\bibliography{references}

\onecolumngrid 
\newpage

\begin{center}
{\bf \large Supplementary Material for ``Ergodicity breaking meets criticality in a gauge-theory quantum simulator''}
\end{center}
\begin{center}
Ana Hudomal$^1$, Aiden Daniel$^2$, Tiago Santiago do Espirito Santo$^3$, Milan Kornja\v ca$^4$,\\ Tommaso Macrì$^4$, Jad C.~Halimeh$^{5,6,7,8}$, Guo-Xian Su$^{9,10}$, Antun Bala\v z$^1$, and Zlatko Papi\'c$^2$\\
\vspace*{0.1cm}
{\footnotesize
$^1$\emph{Institute of Physics Belgrade, University of Belgrade, 11080 Belgrade, Serbia}\\
$^2$\emph{School of Physics and Astronomy, University of Leeds, Leeds LS2 9JT, United Kingdom}\\
$^3$\emph{Department of Theoretical and Experimental Physics, Federal University of Rio Grande do Norte, 59078-970 Natal, RN, Brazil}\\
$^4$\emph{QuEra Computing Inc., 1380 Soldiers Field Road, Boston, MA, 02135, USA}\\
$^5$\emph{Department of Physics and Arnold Sommerfeld Center for Theoretical Physics (ASC), Ludwig Maximilian University of Munich, 80333 Munich, Germany}\\
$^6$\emph{Max Planck Institute of Quantum Optics, 85748 Garching, Germany}\\
$^7$\emph{Munich Center for Quantum Science and Technology (MCQST), 80799 Munich, Germany}\\
$^8$\emph{Department of Physics, College of Science, Kyung Hee University, Seoul 02447, Republic of Korea}\\
$^9$\emph{Department of Physics, Massachusetts Institute of Technology, Cambridge, MA 02139, USA}\\
$^{10}$\emph{MIT-Harvard Center for Ultracold Atoms, Cambridge, MA 02139, USA}
}
\end{center}
\setcounter{section}{0}
\setcounter{subsection}{0}
\setcounter{equation}{0}
\setcounter{figure}{0}
\renewcommand{\theequation}{S\arabic{equation}}
\renewcommand{\thefigure}{S\arabic{figure}}
\renewcommand{\thesection}{S\arabic{section}}
\renewcommand{\thesubsection}{\Alph{subsection}}

\vspace*{0.2cm}

{\footnotesize This supplementary material contains additional details of the experimental setup and measurement procedure, numerical verifications of the encoding of the quantum link model (QLM) onto a Rydberg atom array, and further results about the dynamical phase diagram and the role of electron-positron pairs in ergodicity-breaking dynamics. 
}

\section{Experimental setup and choice of parameters}

Here we provide additional details about the experimental setup and justify  our choice of parameters. All measurements reported in the main text were carried out on QuEra's Aquila device~\cite{AquilaWhitepaper}, based on programmable arrays of Rydberg atoms. As mentioned  in the main text, the Hamiltonian describing a Rydberg atom tweezer array is
\begin{equation}
    \hat{H}_\mathrm{Ryd}=\frac{\Omega}{2} \sum_{j=1}^L \hat{X}_j -\Delta \sum_{j=1}^L \hat{n}_j + \sum_{i<j}^L V_{ij}\hat{n}_i\hat{n}_j\,,
    \label{eq:ryd2}
\end{equation}
where $\hat{X} \equiv \lvert g\rangle\langle r\rvert+\lvert r\rangle\langle g\rvert$ encodes Rabi oscillations between the ground state $|g\rangle$ and excited state $|r\rangle$ with the frequency $\Omega$, $\hat{n} \equiv \lvert r\rangle\langle r\rvert$ counts the Rydberg excitations, $\Delta$ is the detuning, and $V_{ij}=C_6/r^6_{ij}$ is the van der Waals interaction between two atoms in the excited Rydberg states at a distance $r_{ij}$. 

We will focus on 1D arrays in the blockade regime, $V_{i,i+1}\gg \Omega, \Delta$, where neighboring excitations $\ldots rr \ldots$ are energetically unfavorable. This regime is achieved by tuning the Rydberg blockade radius $R_b$, as discussed in Sec.~\ref{sm:agreement}. 

For the optimal choice of $R_b$, the model in Eq.~(\ref{eq:ryd2}) can be mapped to the spin-$1/2$ QLM in the main text if we set $J=\Omega/2$ and $m=\Delta/2$, with the density of Rydberg excitations $\hat n$ mapping to the charge density $\hat Q$ in the QLM~\footnote{\mbox{Note that our Rydberg conventions differ by an overall factor of 2 compared to Ref.~\cite{Surace2020}.}}.
As explained in Appendix~A of the main text, the allowed states of matter and the surrounding gauge fields in the QLM are mapped onto the states of a Rydberg dimer, with the following rules for the even and odd sites in the lattice:
\begin{eqnarray}\label{eq:mapping_rules}
\text{even} &:& {rg}, {gg}, {gr} \leftrightarrow {\triangleright}{\circ}{\triangleright}, {\triangleright}{\color{blue}\bullet}{\blacktriangleleft}, {\blacktriangleleft}{\circ}{\blacktriangleleft}, \\
\text{odd} &:& {rg}, {gg}, {gr} \leftrightarrow {\blacktriangleleft}{\circ}{\blacktriangleleft}, {\blacktriangleleft}{\color{red}\bullet}{\triangleright},  {\triangleright}{\circ}{\triangleright},
\end{eqnarray}
while the terms $rr$ represent gauge violations that should be energetically suppressed. Applying these rules sequentially, the main states of interest in the QLM model---the electric flux and charge-proliferated states---have the following representations in terms of $\mathbb{Z}_2$ and polarized states in the Rydberg atom model:
\begin{equation}\label{eq:states_mapping}
    \lvert{...\triangleright}{\circ}{\triangleright}{\circ}...\rangle  \leftrightarrow |...grgr...\rangle, \lvert{...\blacktriangleleft}{\circ}{\blacktriangleleft}{\circ}...\rangle \leftrightarrow |...rgrg...\rangle, \lvert{...\triangleright}{\textcolor{blue}\bullet}{\blacktriangleleft}{\textcolor{red}\bullet}{\triangleright}{\textcolor{blue}\bullet}...\rangle \leftrightarrow |...gggg...\rangle.
\end{equation}
While the results in the main text were primarily expressed in the QLM picture, in the remainder of this supplementary material, we will focus on the Rydberg notation and corresponding quantities as they are directly measured in our experiments.

\begin{center}
\begin{table}
\begin{tabular}{|l | c|}  
 \hline
 \multicolumn{2}{|c|}{\textbf{Aquila parameters}} \\
 \hline\hline
 Width of the available space $w$ & $75 \ \mu s$  \\ 
 \hline
 Height of the available space $h$ & $120 \ \mu s$  \\ 
 \hline
 Minimal interatomic distance $d_\mathrm{min}$ & $4 \ \mu s$  \\ 
 \hline
 Spatial resolution $\delta d$ & $0.1 \ \mu s$  \\ 
 \hline
 Rabi frequency range $(\Omega_\mathrm{min},\Omega_\mathrm{max})$ & $(0.0, 15.8) \ \mathrm{rad} /\mu s$  \\ 
 \hline
 Maximal Rabi freq. slew rate $\mathrm{d}\Omega/\mathrm{d}t$ & $250 \ \mathrm{rad} /\mu s^2$  \\ 
 \hline
 Detuning range $(\Delta_\mathrm{min},\Delta_\mathrm{max})$ & $(-125, 125) \ \mathrm{rad} /\mu s$ \\
 \hline
 Maximal detuning slew rate $\mathrm{d}\Delta/\mathrm{d}t$ & $2500 \ \mathrm{rad} /\mu s^2$ \\
 \hline
 Interaction coefficient $C_6$ & $5.42 \times 10^ 6 \ \mathrm{rad} . \mu m^6 / \mu s$ \\
 \hline
 Maximal experimental time $t_\mathrm{max}$ & $4 \ \mu s$ \\
 \hline
 Temporal resolution $\delta t_\mathrm{min}$ & $0.01 \ \mu s $ \\
 \hline
 \end{tabular}
 \caption{Technical specifications of the Aquila device \cite{AquilaWhitepaper}.}
 \label{tab:aquila}
 \end{table}
\end{center}

\subsection{Spatial arrangement of atoms, parameter values and number of shots}

The flexibility of the Aquila platform allows to modulate the functional form of the Rabi frequency $\Omega(t)$ and detuning $\Delta(t)$, as well as the spatial arrangement of up to 256 atoms.
The technical specifications of the device are listed in Table~\ref{tab:aquila}. Since the extent of the physical space available for the experiment is limited, we want to make the interatomic distance as small as possible in order to form a chain with the largest possible number of atoms. However, the blockade radius depends on the Rabi frequency. We estimate the minimal blockade radius using the maximal available value of the Rabi frequency in Table~\ref{tab:aquila}:
\begin{equation}
    {R_b}^\mathrm{min} = \left( \frac{C_6}{\Omega_\mathrm{max}}  \right)^{\frac{1}{6}} \approx 8.4 \ \mu m.
\end{equation}
Together with $R_b=1.4a$, this yields the atomic spacing $a = 6 \ \mu m$. As we explain in Sec.~\ref{sm:agreement} below, the value $R_b=1.4a$ was chosen because it provides the best agreement with the QLM model. The corresponding Rabi frequency and the detuning range are given by
\begin{equation}
    \Omega \approx 15.43 \ \mathrm{rad}/\mu s, \quad \Delta / \Omega \in (-8.10, 8.10),
\end{equation}
which amounts to $m / J \in (-8.10, 8.10)$ in the QLM language. 

\begin{figure}[tbh]
\includegraphics[width=0.35\textwidth]{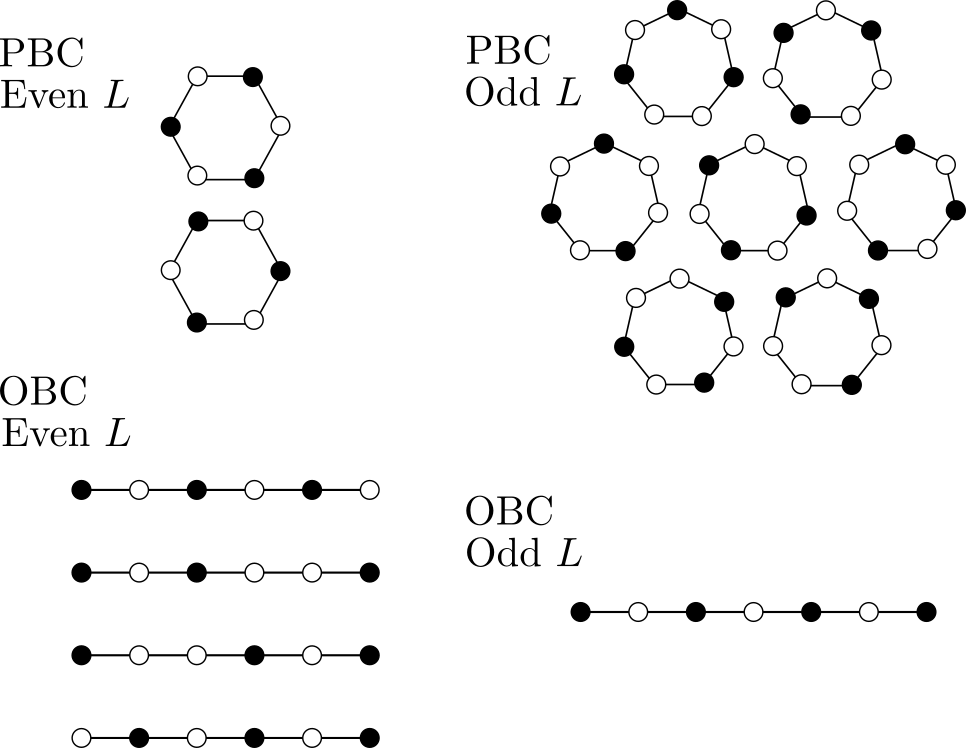}
 \caption{Ground state configurations at $\Delta\rightarrow\infty$ ($m\rightarrow\infty$) for different boundary conditions and system sizes.
 Black circles ($\bullet$) denote Rydberg excitations, while empty circles ($\circ$) denote atoms in the ground state.  }\label{fig:ground_states}
 \end{figure}

The electric flux states map to $\mathbb{Z}_2$-ordered states of Rydberg atoms and they can be strongly affected by boundaries in moderate system sizes.  In Fig.~\ref{fig:ground_states} we show some examples of such configurations, corresponding to the ground states with detuning $\Delta\rightarrow \infty$ (or $m\rightarrow \infty$). The case of periodic boundary conditions (PBCs) and odd system size $L$ is incompatible with $\mathbb{Z}_2$ order, while with open boundary conditions (OBCs) and $L$-even, the ground state has low overlap with the $\mathbb{Z}_2$ states. This is due to the presence of equal-energy configurations with $\ldots gg\ldots$ defects, whose number increases linearly with $L$. Thus, the optimal choices are PBC with $L$-even, with the superposition of two $|\mathbb{Z}_2\rangle$ states as the ground state, and OBC with $L$-odd with a single $|\mathbb{Z}_2\rangle$-type state as the lowest-energy configuration. While we expect the results to remain qualitatively the same for both choices~\cite{Daniel2023}, the OBC case with $L$-odd is more practical for experimental implementation and we adopt it for all the results in the main text.

Based on previous discussion, in Fig.~5 of the main text we showed two types of atom arrangements used in our experiments, both with OBCs. The first  employs parallelization over six rows of 21 atoms each, while the second is a single long chain of 61 atoms.
Since the preparation of long chains comes with a higher error rate and we want to limit the experimental number of shots, we use the short-chain configuration for the majority of measurements and selectively confirm the validity of our results in larger systems by repeating the measurements for a few representative parameter values using the long-chain configuration.

As the typical number of shots needed per timestep is around $100$, when running six chains in parallel it is sufficient to do ${\sim}20$ shots per timestep. However, at least 20-40 timesteps per each point in the phase diagram are needed to simulate the dynamics. The number of required timesteps depends on the oscillation frequency, but it is also limited by the minimal temporal resolution listed in Table~\ref{tab:aquila}. Thus, simulating the full $(\Delta_i,\Delta_f)$ phase diagram with a reasonable resolution (at least $20\times20$) would require a large amount of resources. Because of this, in the main text, we chose three representative line cuts through the phase diagram, one below, one above and one at the critical point.

Finally, the total evolution time is limited to $t_\mathrm{max}= 4 \ \mu s$, which has to accommodate both state preparation and quench dynamics. Our calculations show that $t_\mathrm{ramp}= 1.5 \ \mu s$ is optimal for the ramping protocol and $t_\mathrm{dyn}= 3.5 \ \mu s$ for the quench dynamics. 
At the start of the experiment, the detuning is set to the lowest accessible value $\Delta_0=-8.10\Omega$, for which the ground state is well-approximated by the product state with no Rydberg excitations, corresponding to the charge-proliferated state in the QLM language.
To map out the dynamical phase diagram, we employ a sigmoid-type ramp with an optimal curvature parameter $k=0.88$ and vary the parameters $(\Delta_i,\Delta_f)$. To study the effect of domain walls on the dynamics, we fix $(\Delta_i,\Delta_f)=(2\Omega,0)$ and vary the curvature parameter over the range $k\in[0.00,0.90]$ realizing a broad range of ramp speeds at the critical point.

\FloatBarrier
\subsection{Blockade radius}\label{sm:agreement}

Finding the optimal parameter regime with a good agreement between the QLM and full Rydberg models requires a careful analysis of both ground-state and dynamical properties. We determined the optimal blockade radius $R_b$ based on exact diagonalization studies of the two models, comparing both their ground-state wave functions and the dynamics after the quench. 

\begin{figure}[bt]
 \includegraphics[width=0.35\textwidth]{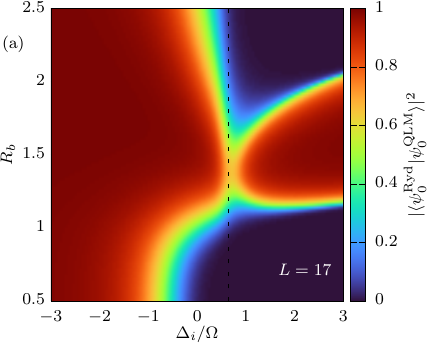}
 \caption{
 The overlap of the ground states of QLM and Rydberg models at different values of detuning $\Delta_i$ and blockade radius $R_b$. The dashed white line at $\Delta_c=0.65\Omega$ marks the phase transition. 
 System size $L=17$. 
 }\label{fig:gs}
\end{figure}

\begin{figure}[tbh]
 \includegraphics[width=0.32\textwidth]{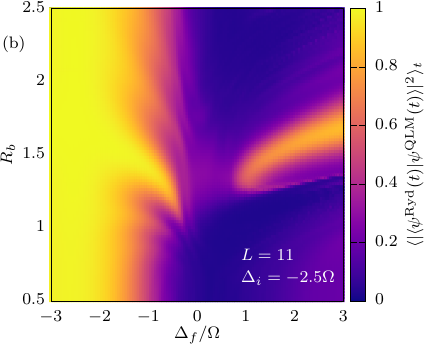}
 \includegraphics[width=0.32\textwidth]{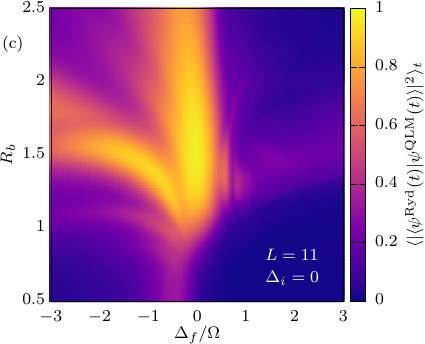}
 \includegraphics[width=0.32\textwidth]{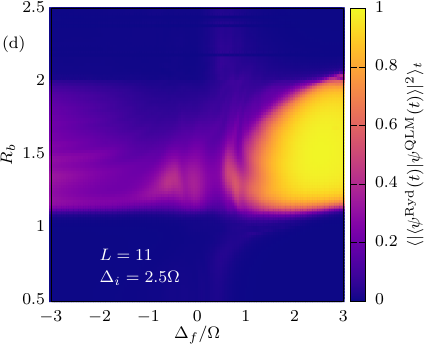}
 \caption{
 Average comoving fidelity starting from the ground states of $H_\mathrm{Ryd}$ and $H_\mathrm{QLM}$ at different $\Delta_i$ and system size $L=11$.
 (a) $\Delta_i/\Omega=-2.5$. (b) $\Delta_i/\Omega=0$. (c) $\Delta_i/\Omega=2.5$.
 }\label{fig:com-fid}
 \end{figure}

Figure~\ref{fig:gs} shows the overlap of the ground states of $\hat{H}_\mathrm{QLM}$ 
and $\hat{H}_\mathrm{Ryd}$ at different $\Delta_i$ and $R_b$. The Hilbert spaces of the two models are not the same because of the Gauss law imposed on the QLM. Nevertheless, the latter is a subset of the full Rydberg Hilbert space, hence we can expand the QLM ground state to the unconstrained Hilbert space and then directly evaluate the inner product between the two states. We see that the best agreement is consistently achieved around $R_b\approx 1.5$, although the overlap is slightly lower around the Coleman critical point, $\Delta_c\approx0.65\Omega$.

In order to compare the dynamics of the QLM with the Rydberg model, we use the comoving fidelity,
\begin{equation}
    F_c(t)=\lvert\langle \psi_\mathrm{QLM}(t)| \psi_\mathrm{Ryd}(t)\rangle\rvert^2,
    \label{eq:com-fid}
\end{equation}
which represents the overlap between the instantaneous wave functions at different times. Similar to the ground state overlap, the fidelity is computed by first expanding the QLM state into the unconstrained Hilbert space. In Fig.~\ref{fig:com-fid} we compute the average comoving fidelity over the time interval $t\in[0,20]$ for the initial ground states of $H_\mathrm{QLM}$ and $H_\mathrm{Ryd}$ at three different values of $\Delta_i$. In contrast to the ground-state overlap, the comoving fidelity has a stronger dependence on detuning as we vary $\Delta_f$. Nevertheless, there is a range of $R_b\approx 1.5a$ where the average comoving fidelity is sufficiently high for all considered values of $\Delta_f$.

\begin{figure}[hbt]
 \includegraphics[width=0.32\textwidth]{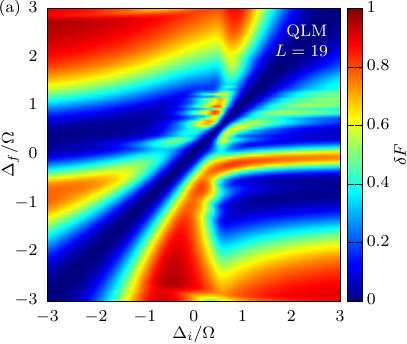}
 \includegraphics[width=0.32\textwidth]{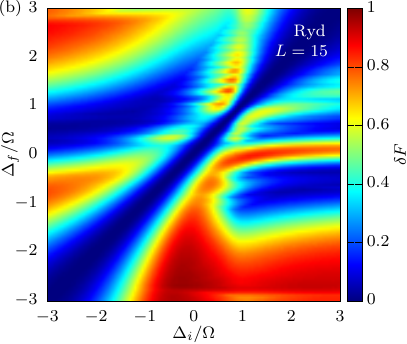}
 \caption{Comparison of the dynamical phase diagram between the QLM and Rydberg models for the blockade radius $R_b=1.4a$. The initial state is the ground state at detuning $\Delta_i$, while the evolution is performed using $\Delta_f$. The color represents the fidelity difference, Eq.~(\ref{eq:deltaf}). The system size is $L=19$ [QLM in (a)] and $L=15$ [Rydberg model in (b)].
 }\label{fig:phase-diag}
 \end{figure}

 Finally, we consider the dynamical phase diagram obtained by preparing the system in the ground state at detuning $\Delta_i$ and then evolving with a quenched detuning $\Delta_i \to \Delta_f$. To detect the ergodicity-breaking regimes associated with quantum many-body scars (QMBSs), we compute the quantum fidelity
\begin{equation}
    F(t)=\lvert\langle \psi(0)| \psi(t)\rangle\rvert^2,
    \label{eq:fidelity}
\end{equation}
and consider the difference between its maximal and minimal values over a time interval:
\begin{equation}\label{eq:deltaf}
    \delta F \equiv \max_{t\in[1,20]} F(t) - \min_{t\in[1,20]} F(t).
\end{equation}
For a generic initial state, the fidelity is expected to drop to zero by time order $t\approx 1$ and remain around that value at subsequent times, hence $\delta F \approx 0$. By contrast, for a QMBS reviving initial state, the fidelity will rise to a value close to 1 at some later time, hence $\delta F$ will be non-zero. Figure~\ref{fig:phase-diag} illustrates the dependence of $\delta F$ across the $(\Delta_i,\Delta_f)$ phase diagram 
for the QLM and full Rydberg models, for the blockade radius $R_b=1.4a$. We see that the main features of the phase diagram, in particular the ergodicity-breaking regimes, are in good agreement between the two models for the chosen value of $R_b$.

\FloatBarrier
\subsection{Phase transition point}\label{sm:phase_transition}

Both the full Rydberg model, Eq.~(\ref{eq:ryd2}), 
and the QLM model contain a second order phase transition which belongs to the Ising universality class. For even system sizes, this marks a transition between a non-degenerate and doubly-degenerate ground states. For odd system sizes, there is a single ground state on either side of the transition, but the gap to the first excited state still closes in the thermodynamic limit. 
Here we determine the phase transition point in the two models by identifying a point where the gap is smallest and extrapolating to the infinite system size limit.

\begin{figure}[bth]
 \includegraphics[width=0.32\textwidth]{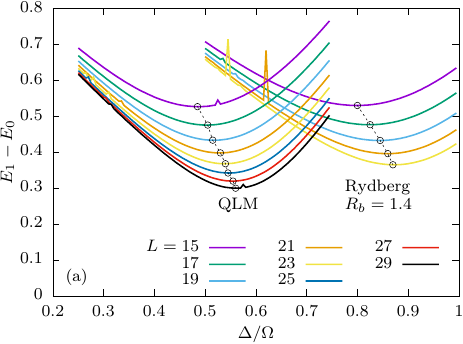}
 \includegraphics[width=0.32\textwidth]{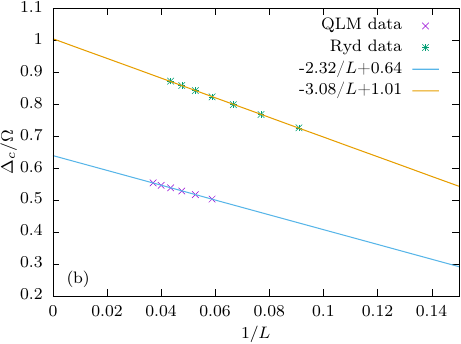}
 \includegraphics[width=0.32\textwidth]{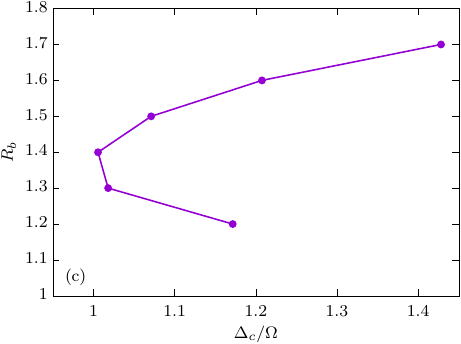}
\caption{
The location of the phase transition.
(a) The gap between the ground state and the first excited state at different detunings $\Delta$ and different system sizes in the QLM and Rydberg models. The blockade radius is $R_b=1.4a$.
 (b) System-size extrapolation of the critical detuning $\Delta_c$ corresponding to the minimal gap in panel (a). 
 (c) Phase transition points in the Rydberg model at different $R_b$, determined using the minimal gap criterion.
 }\label{fig:gap}
 \end{figure}
 
In Fig.~\ref{fig:gap}(a) we plot the energy gap between the lowest two states in the spectrum for different system sizes as we scan the detuning $\Delta$. Upon increasing the system size, the point with the minimal gap in the QLM model moves towards the expected value of $0.65\Omega$~\cite{Rico2014}. However, due to the tail of the van der Waals interactions, the phase transition point in the Rydberg case is significantly larger than the QLM value. In Fig.~\ref{fig:gap}(a), we show the results for $R_b=1.4a$, where we found the best agreement between the two models in Sec.~\ref{sm:agreement}. In Fig.~\ref{fig:gap}(b) we perform the system-size scaling to determine the phase transition point in the thermodynamic size limit. 
Since exact diagonalization is limited to relatively small system sizes, our extrapolated value for the QLM model, $\Delta_c \approx 0.64\Omega$, is in reasonably good agreement with the expected value $0.65\Omega$~\cite{Rico2014}. Moreover, our results for the Rydberg model, $\Delta_c \approx \Omega$, are also in line with the literature \cite{Keesling2019}. Finally, the location of the critical point is sensitive to the Rydberg blockade radius $R_b$ that determines the interaction strength, as illustrated in  Fig.~\ref{fig:gap}(c).

\subsection{Deviation from thermal ensemble}

Our diagnostic of ergodicity breaking in the main text, $D_\mathrm{th}$, was defined as the mean-square deviation of charge density $\langle \hat Q(t) \rangle$, or, equivalently, the Rydberg excitation density $\langle \hat n(t) \rangle$, from the corresponding thermal value $n_\mathrm{th}$.  The latter is given by the Boltzmann-Gibbs expression:
\begin{equation}
    n_\mathrm{th}=\textup{Tr}(\hat{\rho}_\mathrm{th}\hat{n}), \quad     \hat{\rho}_\mathrm{th}=\frac{1}{\mathcal{Z}}e^{-\beta \hat{H}}, \quad \mathcal{Z}=\textup{Tr}(e^{-\beta \hat{H}}).
\end{equation}
Here,  $\beta=1/T$ is the inverse temperature ($k_B=1$), which can be obtained from the condition that the expectation value of the Hamiltonian in the initial state is equal to its expectation value in the ensemble with an appropriate temperature:
\begin{equation}
    \langle \psi(t=0)\lvert \hat{H}\rvert \psi(t=0)\rangle=\textup{Tr}(\hat{\rho}_\mathrm{th}\hat{H}). 
\end{equation}
Since the thermal value of the excitation density $n_\mathrm{th}$ is not easily extracted from the available experimental data, we use other quantities to approximate it. In the ergodic case (for a generic initial state), the system is expected to relax to the thermal value after a sufficiently long time. We can therefore estimate $n_{th}$ by averaging over long time intervals. However, the experimental time window is limited by decoherence and cannot be arbitrarily increased, hence we need to establish an acceptable range for it.

\begin{figure}[tb]
 \includegraphics[width=0.5\textwidth]{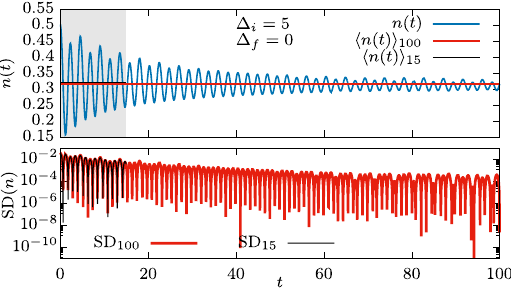}
 \caption{
 (a) The excitation density dynamics for $\Delta_i=5\Omega$ and $\Delta_f=0$. The red line represents the average value of $n(t)$ over the full observed period $t\in[0,100]$, while the black line is the average over the initial period $t\in[0,100]$ (shaded area), which roughly corresponds to the experimentally-accessible timescale.
 (b) Evolution of the square deviation, calculated by approximating the relaxation value with $\langle n(t)\rangle_{100}$ (red) and $\langle n(t)\rangle_{15}$ (black), showing that there is no significant difference between the two. 
 }\label{fig:msd_dynamics}
 \end{figure}

In Fig.~\ref{fig:msd_dynamics} we plot the evolution of $n(t)$ at $(\Delta_i,\Delta_f)=(5\Omega,0)$ and compare the excitation density averaged over the time intervals $t\in[0,100]$ and $t\in[0,15]$. The two intervals yield very similar average values. Moreover, the evolution of the square deviation $\mathrm{SD}(t)$, calculated using $\langle n(t)\rangle_{15}$, does not significantly differ from that computed using $\langle n(t)\rangle_{100}$. This justifies our method of estimating $D_\mathrm{th}$ in the main text.

\FloatBarrier
\subsection{Postprocessing and error mitigation}

Here we provide additional details related to the postprocessing of the experimental data, extracting relevant quantities and performing error-mitigation procedures.

For each data point $(\Delta_i,\Delta_f,k,t_m)$, we perform 40 experimental shots on six parallel chains of $L=21$ atoms, yielding a total of 240 shots, and 200 shots for longer chains with $L=61$ atoms. All runs with detected atom loss due to incorrect initial loading are automatically discarded, leaving approximately 180 valid shots per data point for the short chains and 120 shots for the long chains. 

The single-atom measurement outcomes obtained in this procedure are stored as bitstrings and subsequently processed to extract the relevant observables, including the Rydberg excitation density $\langle \hat n(t) \rangle$, the charge density $\langle \hat Q(t) \rangle$, and other quantities of interest. Unless stated otherwise, error bars correspond to standard errors computed over all bitstrings. Because the evaluation of the PCA entropy $S_\mathrm{PCA}$ requires the singular values of the bitstring data matrix, we estimate its uncertainty using a resampling procedure. Specifically, we generate $n_s=100$ artificial replica datasets by sampling the measured bitstrings with replacement, perform singular value decompositions on each replica, and compute the standard error of the resulting PCA entropy values.

In order to mitigate readout errors, we perform the procedure described in Ref.~\cite{Bravyi2021}. For the Aquila device, the known probabilities of erroneously measuring the state of an atom as ground instead of excited
is $p_{r\rightarrow g} = 8\%$, and excited instead of ground $p_{
g\rightarrow r} = 1\%$~\cite{AquilaWhitepaper}. This amounts to the readout error matrix
\begin{align}
    \hat{A}=\begin{bmatrix}
1-p_{r\rightarrow g} & p_{g\rightarrow r} \\
p_{r \rightarrow g} & 1-p_{g\rightarrow r}
\end{bmatrix}
=\begin{bmatrix}
0.92 & 0.01 \\
0.08 & 0.99
\end{bmatrix},
\end{align}
which is the same for all atoms. For an observable which can be written in a tensor product form $\hat{O}=\hat{O}_1\otimes\hat{O}_2\otimes...\otimes\hat{O}_L$, its error-mitigated expectation value can be computed as
\begin{equation}
    O_\mathrm{EM}=\frac{1}{n}\sum_i^n\prod_j^L\langle e\rvert \hat{O}_j\hat{A}^{-1}\lvert s_j^i\rangle,
\end{equation}
where $\lvert e\rangle=\lvert g\rangle+\lvert r \rangle$ is a single-atom state and $s_j^i\in\{g, r\}$ is the measurement outcome for the $j$-th bit of the $i$-th bitstring $s^i\in\{g, r\}^L$ of out $n$ viable shots. This procedure results in an increased excitation density $\langle \hat n_\mathrm{EM}(t) \rangle$, and consequently, reduced charge density $\langle \hat Q_\mathrm{EM}(t) \rangle$.

When applicable, we also employ additional procedures to mitigate the effects of readout errors. In particular, in measurements of gauge violations, identified as neighboring Rydberg excitations of the form $\lvert...r r...\rangle$, we find that configurations of the type $\lvert...rrr...\rangle$ contribute disproportionately to this quantity, with a much larger weight than the other configurations with the same number of violations, such as $\lvert...rr...rr...\rangle$. Given that the Rydberg blockade radius is $R_b=1.4a$, nearest-neighbor excitations are strongly suppressed, implying that three consecutive excitations are extremely unlikely and are therefore most plausibly attributed to a readout error on the central atom. A simple substitution $\lvert rrr\rangle\rightarrow\lvert r g r\rangle$ significantly improves the agreement between experimental measurements and numerical predictions, as shown in Fig.~3(h) of the main text.

\section{Additional experimental data}\label{a:experimental_data}

In this Section we provide additional experimental results, including the $Q$ and $S_\mathrm{PCA}$ dynamics, their Fourier transforms, and the statistical analysis of experimental bitstrings.

\subsection{Quench dynamics}

\begin{figure}[tbh]
 \includegraphics[width=0.99\textwidth]{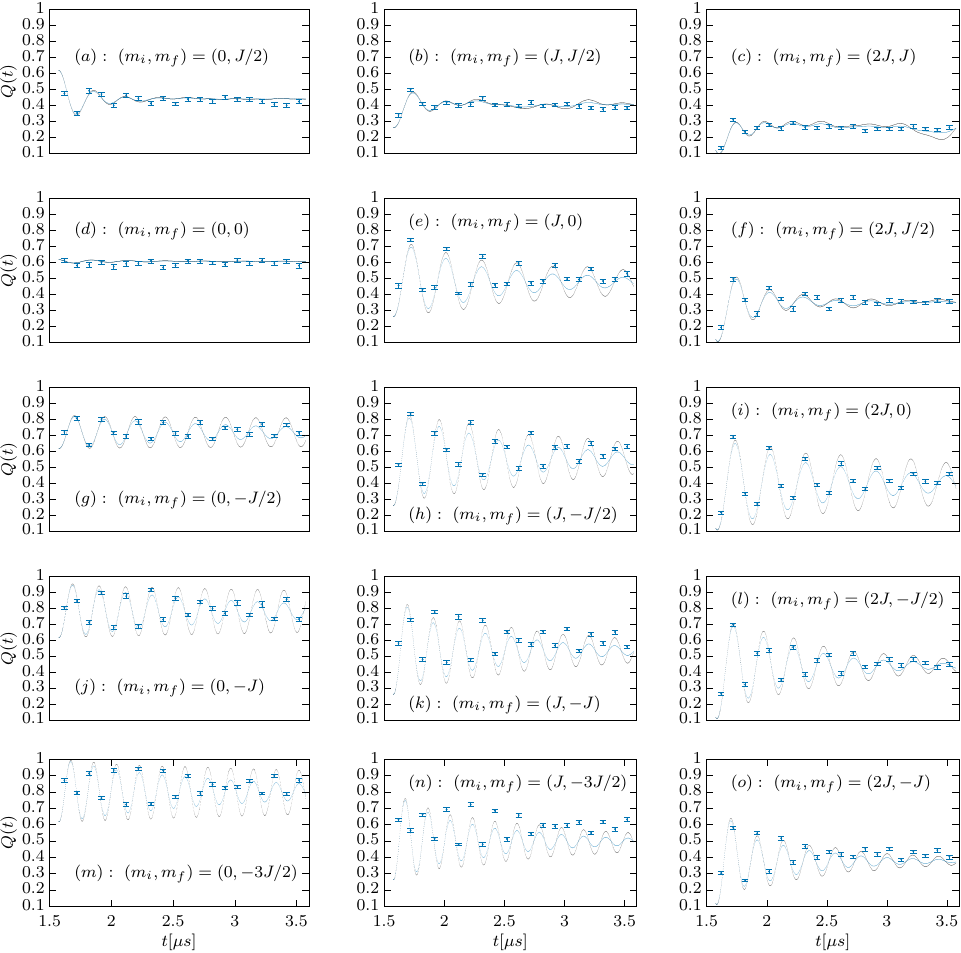}
 \caption{Dynamics of charge density $Q(t)$ for different $(m_i,m_f)$ points along the three cuts through the phase diagram. The points are experimental measurements for $L=61$, while the blue and black solid lines are numerical results for $L=21$, with and without damping, respectively.
 }\label{fig:q_dynamics}
 \end{figure}

\begin{figure}[tbh]
 \includegraphics[width=0.99\textwidth]{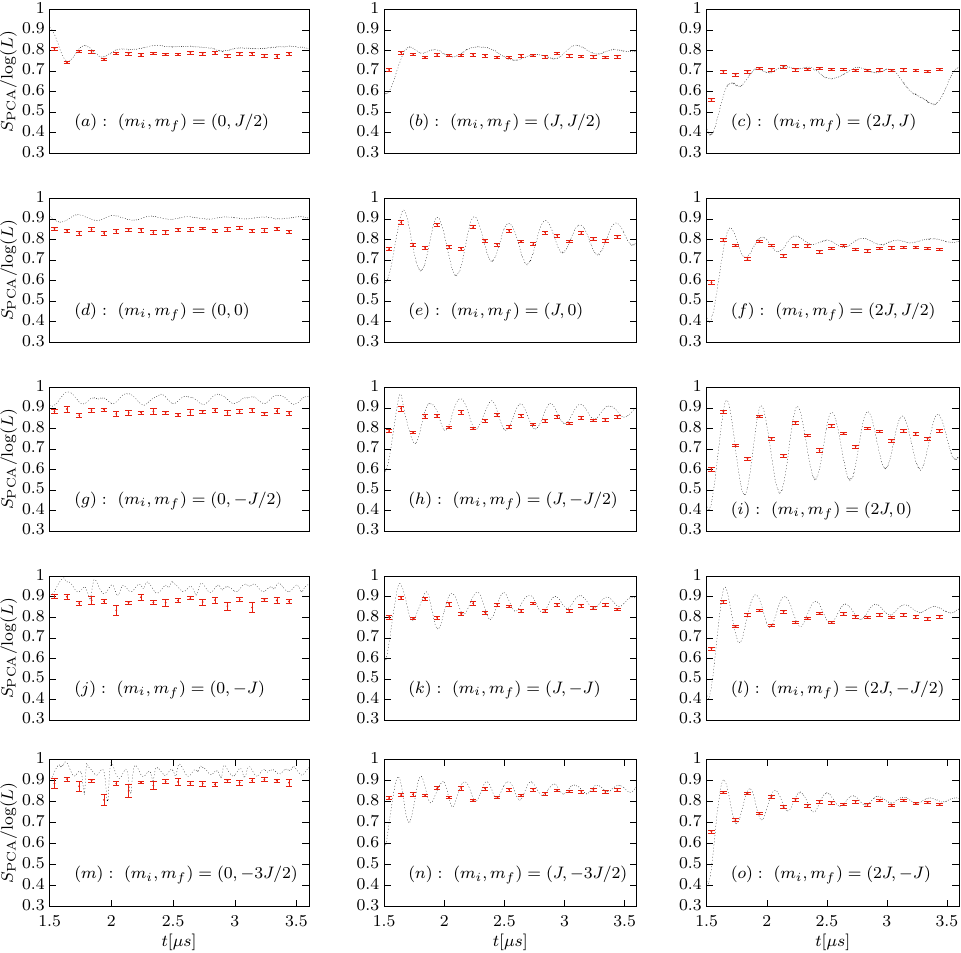}
 \caption{Dynamics of normalized PCA entropy $S_\mathrm{PCA}(t)/\log(L)$, for different $(m_i,m_f)$ points along the three cuts through the phase diagram (same parameters as in Fig.~\ref{fig:q_dynamics}). The points are experimental measurements for $L=61$, while the black and blue solid lines are numerical results for $L=21$ with $n_s=10000$ samples.
 }\label{fig:spca_dynamics}
 \end{figure}

In Fig.~2 of the main text, we presented time series plots of charge density $Q(t)$ and PCA entropy $S_\mathrm{PCA}(t)$ for two illustrative points in the phase diagram. Figure~\ref{fig:q_dynamics} shows more systematically the evolution of charge density $Q(t)$ after the quench $m_i\rightarrow m_f$ in the QLM. The corresponding results for the PCA entropy $S_\mathrm{PCA}(t)$ are shown in Fig.~\ref{fig:spca_dynamics}. Both quantities display good agreement between the experimental data and the numerical prediction. In particular, the experiment is clearly able to distinguish between ergodic and non-ergodic regimes, with the same frequency of scar-induced oscillations in both measured quantities. We note that the agreement between theory and experiment in the case of PCA entropy is notably worse in some cases compared to the charge density, e.g., in the benchmark case $m_i=m_f=0$ where the initial state is close to an eigenstate of the quench Hamiltonian (for sufficiently-accurate state preparation), hence no dynamics is expected. We attribute this discrepancy to the readout errors which, as explained in Appendix~B of the main text, were not mitigated in the case of PCA entropy. By contrast, the (error-mitigated) result for $Q(t)$ is in good agreement with theory.  

\begin{figure}[tbh]
  \includegraphics[width=0.95\textwidth]{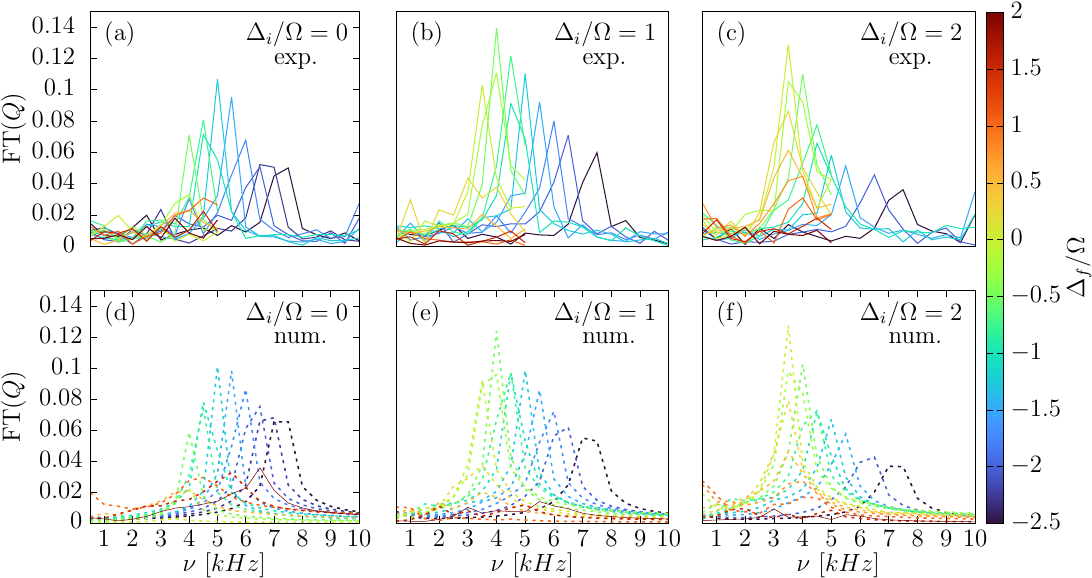}
 \caption{
 Fourier transform of $\langle Q(t)\rangle$ for the system size $L=21$.
 Top: experiment. Bottom: numerics. Left: $\Delta_i=0$. 
 Center: $\Delta_i=\Omega$. Right: $\Delta_i=2\Omega$.
 }\label{fig:fourier_transform}
 \end{figure}

To quantify the quality of the revivals and identify the dominant frequency, we compute the Fourier transform of the charge density $\langle Q(t)\rangle$ and present the results along the three experimental cuts $\Delta_i/\Omega\in\{0,1,2\}$ in Fig.~\ref{fig:fourier_transform}. Different colors correspond to different post-quench detunings $\Delta_f$, which govern the dynamics. The solid lines in the top row display the experimental data [(a)-(c)], while the dashed lines in the bottom row show the results of numerical simulations for $L=21$ [(d)-(f)]. For all combinations of $(\Delta_i,\Delta_f)$, we find good quantitative and excellent qualitative agreement in all cases, in particular for the value of $\Delta_f$ that produces the highest peak, signifying the most prominent QMBS regime for a given $\Delta_i$. For all three values of $\Delta_f$, the peak is barely visible for $\Delta_f>0.5\Omega$, meaning that those parameters are outside of the QMBS regime, while for $\Delta_f<0$ the highest peak moves towards higher frequencies with decreasing $\Delta_f$. In cases where the highest peak was outside the initial frequency range, we performed additional measurements at intermediate time steps to increase the range and correctly locate the peak.

\begin{figure}[tbh]
 \includegraphics[width=0.63\textwidth]{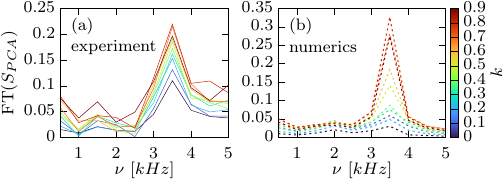}
 \caption{
 Fourier transform of $S_\mathrm{PCA}(t)$ for the system size $L=21$, $\Delta_i=2\Omega$ and $\Delta_f=0$.
 (a) Experiment. 
 (b) Numerics with $n_s=10000$ samples.
 }\label{fig:fourier_spca_k}
 \end{figure}

In Fig.~\ref{fig:fourier_spca_k} we fix the parameters to $(\Delta_i,\Delta_f)=(2\Omega,0)$, where we find strong scar-induced oscillations, and vary the curvature parameter $k$ of the sigmoid-type ramp. Instead of the excitation density, here we perform the Fourier transform of the PCA entropy $S_\mathrm{PCA}(t)$, since this quantity is more sensitive and allows for better differentiation by $k$ values. Once again, we find good agreement with the numerical prediction. The height of the Fourier peak monotonically increases with $k$ until a value around $k\approx 0.85$ is reached, when the ramp becomes too steep at the beginning and the end, and too flat in the critical region.

\FloatBarrier

\FloatBarrier

\subsection{Statistical analysis of experimental bitstrings}\label{a:bitstring_statistics} 

 Here we perform a statistical analysis of experimental bitstrings of length $L=21$, which were used for the study of Kibble-Zurek effect in the main text. The detuning after the ramp and before the quench was set to $\Delta_i=2\Omega$, such that the prepared state resembles one of the $|\mathbb{Z}_2\rangle$ states with defects in the form of consecutive ground-state atoms, or in the QLM language, a state with uniform electric field domains, separated by domain walls formed by a few (anti)particles. Defects in the form of neighbouring Rydberg excitations represent gauge violations and are strongly suppressed due to the Rydberg blockade effect, see Fig.~3(h) in the main text.
 
We use the following definition of different types of domains.
The ground state domains consist of at least two consecutive non-excited atoms, and can be interpreted as strings of (anti)particles.
The $\mathbb{Z}_2$ domains always begin and end with an excitation, unless they are on a boundary of the system. In the QLM interpretation, these domains correspond to vacuum states with an uniform electric field.
For example, the configuration $ \textcolor{blue}{grgrgr}\lvert\textcolor{red}{ggg}\lvert\textcolor{blue}{rgrgr}\lvert\textcolor{red}{gg}\lvert\textcolor{blue}{rgr}\lvert\textcolor{red}{gggg}$ has five domain walls (denoted by vertical lines) between three ${\mathbb{Z}}_2$ (blue) and three ground-state domains (red). 

\begin{figure}[bht]
 \includegraphics[width=0.96\textwidth]{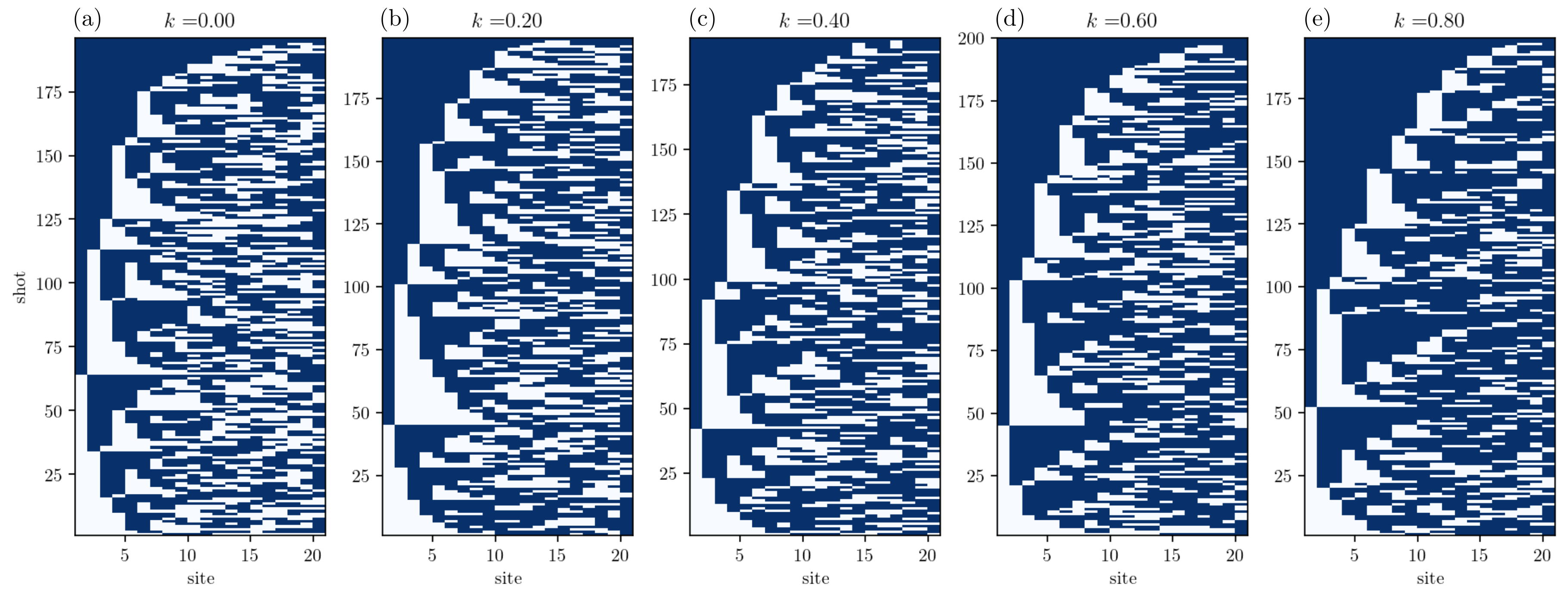} 
 \includegraphics[width=0.96\textwidth]{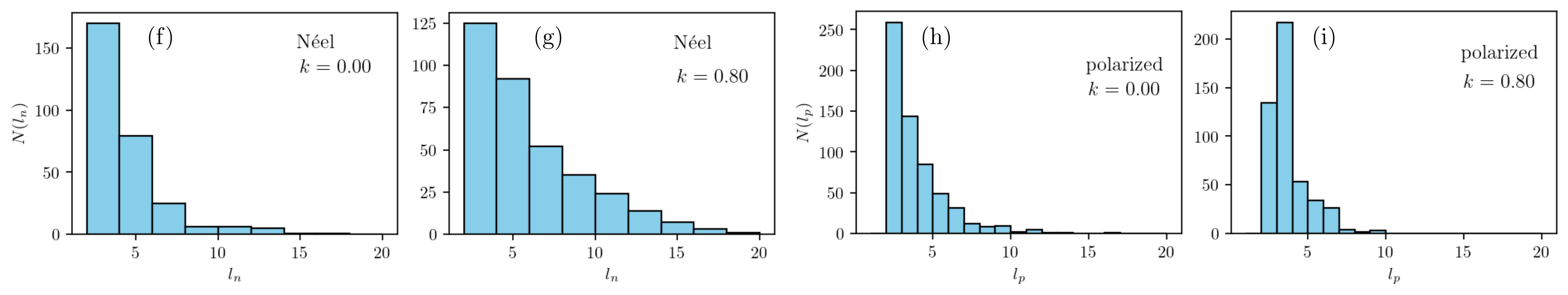}
 \caption{ Bistring statistics.
 (a)-(e) Domains in experimental bitstrings, sorted by the longest $\mathbb{Z}_2$ domain. The $\mathbb{Z}_2$ (N\'eel) domains are shown in blue, while the ground state (polarized) domains and Rydberg-blockade violations are shown in white.
 (a) $k=0$,
 (b) $k=0.2$,
 (c) $k=0.4$,
 (d) $k=0.6$,
 (e) $k=0.8$.
 (f)-(g) Histogram of the length of $\mathbb{Z}_2$ (vacuum) domains, for $k=0$ and $k=0.8$.
 (h)-(i) Histogram of the length of polarized (charge-proliferated) domains, for $k=0$ and  $k=0.8$.
}\label{fig:bitstring_statistics}
 \end{figure}

 Figures~\ref{fig:bitstring_statistics}(a)-(e) show the $\mathbb{Z}_2$ (blue) and polarized (white) domains for states prepared by a sigmoid-type ramp with the curvature parameter in the range $k\in[0,0.8]$. The Rydberg-blockade/gauge violations, whose density is very low, are also represented here in white, since they are another type of defects. The bitstrings are sorted by the length of the longest $\mathbb{Z}_2$ domain. It is clear from Figs.~\ref{fig:bitstring_statistics}(a)-(c) that the bitstrings prepared by slower ramps contain longer antiferromagnetically-ordered domains, with the number of defectless $\mathbb{Z}_2$ states growing with $k$. This is further confirmed by the histograms for the length of the $\mathbb{Z}_2$ domains (note that their lengths are odd numbers due to the definition of domains we use here), see Figs.~\ref{fig:bitstring_statistics}(f)-(g), which show that the numbers of longer domains indeed grow with $k$. Domains of length $1$ are not considered here since they would simply represent an isolated Rydberg excitation. 

Additionally, we plot the histograms for the polarized state domains in Figs.~\ref{fig:bitstring_statistics}(h)-(i). Those have a minimal length $2$, since a single ground-state atom can be considered a part of the neighboring $\mathbb{Z}_2$ domain. We see that the number of such domains decreases, while the most probable ground state domain length for $k=0.8$ changes from $2$ to $3$, which has significant effects on the post-quench dynamics. Numerical simulations of product states with different types of defects are presented in the following Sec.~\ref{sm:product_states_defects}. 

\begin{figure}[tbh]
 \includegraphics[width=0.65\textwidth]{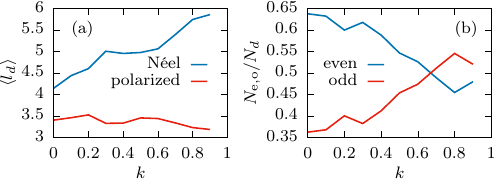}
 \caption{
 Domain length statistics of the experimental data for $L=21$ atoms.
 (a) Average length of $\mathbb{Z}_2$ (vacuum) state and polarized (charge-proliferated) state domains.
 (b) Proportion of even- and odd-length polarized state domains, or respectively, odd- and even-legnth (anti)particle strings.
 }\label{fig:domain_ratio}
 \end{figure}

The bistring statistics findings are summarized in Fig.~\ref{fig:domain_ratio}(a), which shows that the length of the ground-state domains remains approximately the same while the length of $\mathbb{Z}_2$ domains steadily grows with increasing $k$. However, the total number of defects decreases and the proportion of odd-length compared to even-length defects also increases with $k$, as can be seen in Fig.~\ref{fig:domain_ratio}(b). Note that odd (even) length ground state domains correspond to even (odd) strings of (anti)particles.
This finding is somewhat surprising, given that in the LGT interpretation this means that the electric fluxes associated with $\mathbb{Z}_2$ domains tend to be aligned in the same direction across different domains when the ramp is sufficiently slow in the vicinity of the phase transition, implying some correlation between the domains. This effect, which was observed both in the numerical and experimental data, is not explained by the conventional Kibble-Zurek theory.

 \FloatBarrier
 \section{Impact of domain walls on the dynamics}\label{sm:product_states_defects}

 \begin{figure}[tb]
 \includegraphics[width=0.49\textwidth]{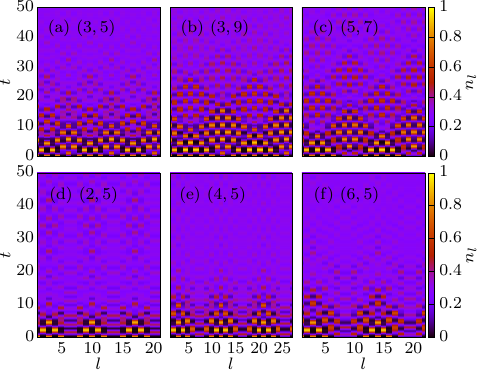}
 \caption{
 Numerical study of the impact of defects on the dynamics in a system size $L=21$. Density of excitations starting from product states with different unit cells $(l_p,l_n)$ (see text).
 (a) $(3,5)$. (b) $(3,9)$. (c) $(5,7)$.
 (d) $(2,5)$. (e) $(4,5)$. (f) $(6,5)$.
 }\label{fig:product_states_defects}
 \end{figure}

In this section, we numerically investigate the impact of different types of defects in the initial state on the dynamics following the quench. For simplicity, the detuning is set to $\Delta=0$ and we work in the effective QLM model, as its constrained Hilbert space allows us to explore larger system sizes. We study the excitation dynamics of several product states formed by periodic repetition of unit cells. Each unit cell consists of a polarized state domain of length $l_p$ and $\mathbb{Z}_2$ domain of length $l_n$, and is denoted by $(l_p,l_n)$. In Fig.~\ref{fig:product_states_defects}, we show the dynamics starting from product states with odd $l_p$ in the top row (a)-(c) and even $l_p$ in the bottom row (d)-(e). 

The difference between these two cases is striking. The states with odd $l_p$ continue to oscillate between the initial state and its translated partner, resembling the scarred dynamics of the $\mathbb{Z}_2$ and anti-$\mathbb{Z}_2$ states. We have confirmed that the overlap of the initial product state with the eigenstates of the QLM model reveals a familiar pattern of QMBS towers~\cite{Turner2017}, hinting that such states with periodically spaced odd-length defects are also scarred. In the LGT framework, these states consist of domains with uniform electric field, separated by adjacent electron-positron pairs, which then move apart until they collide with their other neighbors and scatter back.

On the other hand, oscillations from states with even $l_p$ quickly decay. As can be seen in Figs.~\ref{fig:product_states_defects}(e) and (f), the $\mathbb{Z}_2$ state domains grow until a $...gg...$ pattern is left between them, which cannot be removed due to frustration from the incompatible $\mathbb{Z}_2$ domains. The two-site polarized domain remains frozen and acts as a barrier between smaller $\mathbb{Z}_2$ domains that continue to oscillate but suffer from early-onset finite size effects, which rapidly destroy the oscillations. This case can be though of single electron/positron particles that are unable to propagate in space due to being stuck between opposite electric field domains which periodically keep changing the direction. 

In cases of product states with a mix of even and odd defects, a combination of previously described behavior is seen (data not shown). Parts of the system with odd defects oscillate together, whereas even defects remain as barriers that separate opposite N\'eel domains. Furthermore, states with randomly spaced odd-length defects exhibit less regular oscillations than those with regularly spaced defects as in Figs.~\ref{fig:product_states_defects}(a)-(c). Finally, the experimentally prepared states at $\Delta_i>\Delta_c$ are entangled superpositions of product states with different types of defects between for $\mathbb{Z}_2$ state domains. As demonstrated earlier for both numerical and experimental data, the expectation values of the numbers of even- and odd-length defects strongly depend on the curvature parameter $k$ of the sigmoid ramp. Slower ramps produce a smaller total number of defects, but also a larger proportion of odd-length defects, resulting in better quality of oscillations after the quench.  Periodic driving with optimal driving parameters can completely remove odd-length defects and restore nearly-perfect $\mathbb{Z}_2$ order~\cite{Hudomal2022Driven}.

\end{document}